\documentclass[fleqn,usenatbib]{mnras}

\usepackage[T1]{fontenc}

\DeclareRobustCommand{\VAN}[3]{#2}
\let\VANthebibliography\thebibliography
\def\thebibliography{\DeclareRobustCommand{\VAN}[3]{##3}\VANthebibliography}

\usepackage{graphicx}	
\usepackage{amsmath}	
\usepackage{amssymb}	

\newcommand{\hii}{H\,{\scshape ii}}
\newcommand{\hiir}{H\,{\scshape ii}~region}
\newcommand{\hiirs}{H\,{\scshape ii}~regions}

\title[\hii\ region IRAS 17149$-$3916]{Galactic {\hiir} IRAS 17149$-$3916 - A multiwavelength study}

\author[Potdar et al.]{Ajay Potdar$^1$\thanks{Present affiliation: Space Physics Laboratory, Vikram Sarabhai Space Centre, Thiruvananthapuram 695 022, Kerala, India}, Swagat R Das$^2$\thanks{Corresponding author: dasswagat77@gmail.com (SRD)}, Namitha Issac$^3$, Anandmayee Tej$^1$\thanks{Corresponding author: tej@iist.ac.in (AT)}, Sarita Vig$^1$, C. H. Ishwara Chandra$^4$\\
$^1$Indian Institute of Space Science and Technology, Thiruvananthapuram 695 547, Kerala, India\\
$^2$Indian Institute of Science Education and Research (IISER) Tirupati,  Tirupati 517507, India \\
$^3$Indian Institute of Astrophysics, Koramangala II Block, Bangalore 560 034, India \\
$^4$National Centre for Radio Astrophysics, Tata Institute of Fundamental Research, Pune University Campus, Pune 411007, India\\
}

\date{Accepted XXX. Received YYY; in original form ZZZ}

\pubyear{2020}

\begin{document}
\label{firstpage}
\pagerange{\pageref{firstpage}--\pageref{lastpage}}
\maketitle

\begin{abstract}
This paper presents a multiwavelength investigation of the Galactic {\hiir} IRAS 17149$-$3916. Using the Giant Meterwave Radio Telescope, India, first low-frequency radio continuum observations at 610 and 1280~MHz for this region are presented. The ionized gas emission displays an interesting cometary morphology which is likely powered by the early type source, E4 (IRS-1). The origin of the cometary morphology is discussed under the framework of the widely accepted bow shock, champagne flow, and clumpy cloud mechanisms. The mid- and far-infrared data from {\it Spitzer}-GLIMPSE and {\it Herschel}-Hi-GAL reveal a complex network of pillars, clumps, bubble, filaments, and arcs suggesting the profound influence of massive stars on the surrounding medium. Triggered star formation at the tip of an observed pillar structure is reported. High-resolution ALMA continuum data show a string of cores detected within the identified clumps. The core masses are well explained by thermal Jeans fragmentation and support the hierarchical fragmentation scenario. Four `super-Jeans' cores are identified which, at the resolution of the present data set, are suitable candidates to form high-mass stars.   
\end{abstract}

\begin{keywords}
stars: formation - H II region - ISM: individual objects (IRAS 17149$-$3916)-
radio continuum: ISM
\end{keywords}



\section{Introduction}
{\hiirs}, that are an outcome of the photoionization of newly forming high-mass stars ($\textup{M} \gtrsim 8~M_{\odot}$), not only play a crucial role in understanding processes involved in high-mass star formation but also reveal the various feedback mechanisms at play on the surrounding ambient interstellar medium (ISM) and the natal molecular cloud. Numerous observational and theoretical studies of {\hiirs} have been carried out in the last two decades. However, dedicated multiwavelength studies of star-forming complexes add to the valuable observational database that provide a detailed and often crucial insight into the intricacies involved.
 
In this work, we study the massive star-forming region IRAS 17149$-$3916. This region is named RCW 121 in the catalog of $\rm H{\alpha}$ emission in the Southern Milky Way \citep{1960MNRAS.121..103R}. The mid-infrared (MIR) dust bubble, S6, from the catalog of \citet{2006ApJ...649..759C} is seen to be associated with this complex. IRAS 17149$-$3916 has a bolometric luminosity of $\sim~9.4 \times 10^4~L_{\odot}$ \citep{2006A&A...447..221B}. In literature, several kinematic distance estimates are found for this complex. The near and far kinematic distance estimates range between 1.6 -- 2.2 and 14.5 -- 17.7~kpc, respectively \citep{{1997MNRAS.291..261W},{2004ApJS..154..553S},{2006A&A...447..221B},{2010ApJ...716.1478W}}. In a recent paper, \citet{2014MNRAS.437..606T} use the spectral classification of the candidate ionizing star along with near-infrared (NIR) photometry to place this complex at 2~kpc. This is in agreement with the near kinematic distance estimates and conforms to the argument of  \citet{2006A&A...447..221B} for assuming the near kinematic distance of 2.1~kpc based on the height above the Galactic plane. Based on the above discussion, we assume a distance of 2.0~kpc in this work. 

This star-forming region has been observed as part of several radio continuum surveys at 2.65~GHz \citep{1969AuJPA..11...27B}, 4.85~GHz \citep{1994ApJS...91..111W}, and more recently at 18 and 22.8~GHz by \citet{2013A&A...550A..21S}. Using NIR data, \citet{2006AJ....131..951R} detect a cluster of young massive stars associated with this IRAS source. These authors also suggest IRS-1, the brightest source in the cluster, to be the likely ionizing star of the {\hiir} detected in radio wavelengths. 
\citet{2008A&A...486..807A} probed the $^{12}\rm CO$ molecular gas in the region. Based on this observation, they conclude that RCW 121 and RCW 122 are possibly unrelated star-forming regions belonging to a giant molecular complex while negating the speculation of these being triggered by Wolf-Rayet stars located in the HM~1 cluster. In the most recent work on this source, \citet{2014MNRAS.437..606T} re-visit the cluster detected by \citet{2006AJ....131..951R}. 
These authors also detect three bright {\it Herschel} clumps, the positions of which coincide with three of the 1.2~mm clumps of \citet{2006A&A...447..221B}.    

\begin{figure*}
\centering
\includegraphics[scale=0.5]{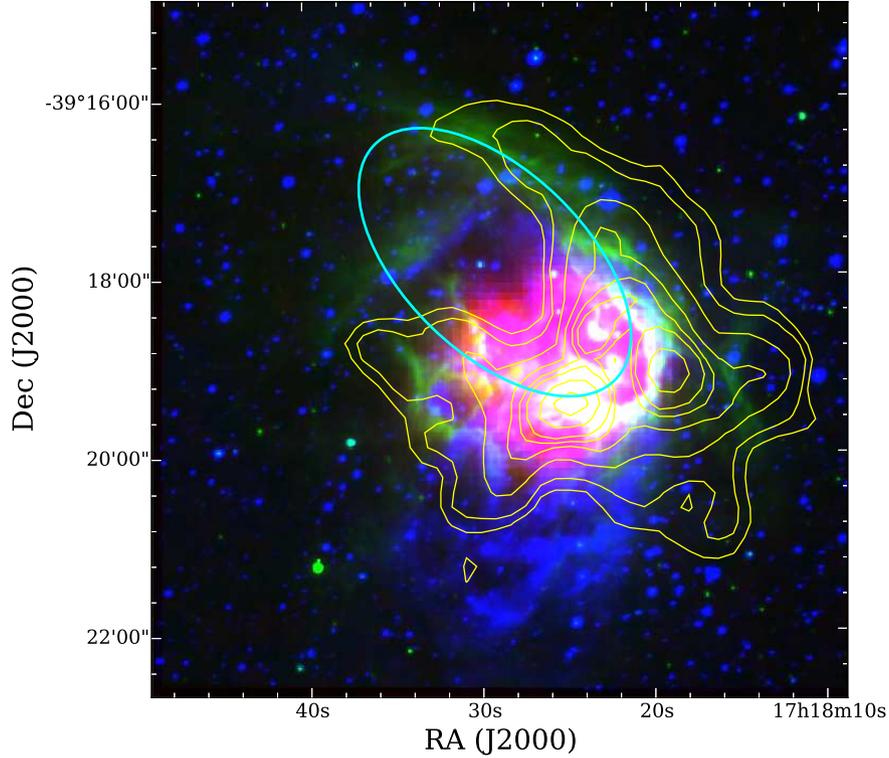}
\caption{Colour-composite image towards IRAS 17149$-$3916 with the MSX 21\,$\rm \mu m$ (red), SuperCosmos $\rm H_{\alpha}$ (blue) and IRAC 5.8 \,$\rm \mu m$ (green) are shown overlaid with the contours of the  \textit{Herschel} 350\,$\rm \mu m$ map. The contour levels are 600, 700, 1000, 1500, 2000, 2500, 4000, and 6000~MJy/sr. The ellipse shows the position and the extent of the bubble, S6, as identified by \citet{2006ApJ...649..759C}.}
\label{fig_intro}
\end{figure*}

Introducing the IRAS 17149$-$3916 complex, in Fig.~\ref{fig_intro}, we show the colour composite image of the associated region. The 5.8\,$\rm \mu m$ IRAC band, which is mostly a dust tracer \citep{2010ApJ...716.1478W}, displays an almost closed, elliptical ring emission morphology. The extent of the bubble, S6, as estimated by \citet{2006ApJ...649..759C} traces this. The cold dust component, revealed by the \textit{Herschel} 350\,$\rm \mu m$ emission, is distributed along the bubble periphery with easily discernible cold dust clumps. Ionized gas sampled in the SuperCosmos $\rm H{\alpha}$ survey \citep{2005MNRAS.362..689P} fills the south-west part of the bubble and extends beyond towards south. MIR emission at 21\,$\rm \mu m$ is localized towards the south-west rim of the bubble. This emission is seen to spatially correlate with the central, bright region of ionized gas (see Fig.~\ref{radioim})
and is generally believed to be due to the Ly$\alpha$ heating of dust grains to temperatures of around 100~K \citep{1991MNRAS.251..584H}.

In this paper, we present an in-depth multiwavelength study of this star-forming region. In discussing the investigation carried out, we present the radio observations and the related data reduction procedure followed in Section \ref{obs-data}. This section also briefly discusses the salient features of the archival data used in the study. Section \ref{results} presents the results obtained for the associated ionized gas and dust environment. Section \ref{discussion} delves into the detailed discussion and interpretation related to the observed morphology of the ionized gas, investigation of the pillar structures, dust clumps in the realm of triggered star formation, and the nature of the detected dust clumps and cores. Section \ref{conclusion} highlights the main results obtained in this study. 
 
\section{Observation, data reduction and archival data} \label{obs-data}
\subsection{Radio Continuum Observation } \label{radio_obs}
To probe the ionized component of IRAS 17149$-$3916, we have carried out low-frequency radio continuum observations of the region at 610 and 1280~MHz with the Giant Meterwave Radio Telescope (GMRT), Pune, India. GMRT offers a hybrid configuration of 30 antennas (each of diameter 45~m) arranged in a Y-shaped layout. The three arms contain 18 evenly placed antennas and provide the largest baseline of $\sim 25$~km. The central $\rm 1\,km^2$ region houses 12 antennas that are randomly oriented with shortest possible baseline of $\sim 100$\,m.
A comprehensive overview of GMRT systems can be found in \citet{1991ASPC...19..376S}.
The target was observed with the full array for nearly full-synthesis to maximize the {\it uv} coverage which is required to detect the extended, diffuse emission. Observations were carried out during August 2014 at 610 and 1280\,MHz with a bandwidth of 32 MHz over 256 channels. In the full-array mode, the resolution is $\sim$5 and 2~arcsec and the largest detectable spatial scale is $\sim$17 and 7~arcmin at 610 and 1280~MHz, respectively. Radio sources 3C 286 and 3C 48 were selected as the primary flux calibrators. The phase calibrator, 1714-252, was observed after each 40-min scan of the target to calibrate the phase and amplitude variations over the full observing run. The details of the GMRT radio observations and constructed radio continuum maps are listed in Table \ref{radio_tab}.

Astronomical Image Processing System (AIPS) was used to reduce the radio continuum data where we follow the procedure detailed in \citet{2017MNRAS.472.4750D} and \citet{2019MNRAS.485.1775I}. The data sets are carefully examined to identify bad data (non-working antennas, bad baselines, RFI, etc.), employing the tasks, {\tt TVFLG} and {\tt UVPLT}. Following standard procedure, the gain and bandpass calibration is carried out after flagging bad data. Subsequent to bandpass calibration, channel averaging is done while keeping bandwidth smearing negligible. Continuum maps at both frequencies are generated using the task {\tt IMAGR}, adopting wide-field imaging technique to account for w-term effects. Several iterations of self-calibration (phase-only) are performed to minimize phase errors and improve the image quality. Subsequently, primary beam correction was carried out for all the generated maps. 

In order to obtain a reliable estimate of the flux density, contribution from the Galactic diffuse emission needs to be accounted for. This emission follows a power-law spectrum with a steep negative spectral index of $-2.55$ \citep{1999A&AS..137....7R} and hence has a significant contribution at the low GMRT frequencies (especially at 610~MHz). This results in the increase in system temperature, which becomes particularly crucial when observing close to the Galactic plane as is the case with our target, IRAS 17149$-$3916. The flux calibrators lie away from the Galactic plane and for such sources at high Galactic latitudes, the Galactic diffuse emission would be negligible. This makes it essential to quantify the system temperature correction to be applied in order to get an accurate estimate of the flux density. Since measurement of the variation in the system temperature of the antennas at GMRT are not automatically implemented during observations, we adopt the commonly used Haslam approximation discussed in \citet{2015MNRAS.451...59M} and implemented in \citet{2019MNRAS.485.1775I}.

The sky temperature, ${T_{\rm sky}}$, at frequency $\nu$ for the location of our source is determined using the equation
\begin{equation}
T_{\rm sky,\nu} = T_{\rm sky}^{408}\bigg(\frac{\nu}{408~\textrm{MHz}}\bigg)^\gamma
\end{equation}
\noindent
where $\gamma = -2.55$ is the spectral index of the Galactic diffuse emission and $\it{T_{\rm sky}^{\rm 408}}$ 
is the sky temperature at 408~MHz obtained from the all-sky 408~MHz survey of \citet{1982A&AS...47....1H}. Using this method, we estimate the scaling factors of 2.2 and 1.24 at 610 and 1280~MHz, respectively, which are used to rescale and obtain the final maps.
\begin{table}
\caption{Details of the GMRT radio continuum observations.} 
\begin{center}
\label{radio_tab}
\begin{tabular}{lll}
\\ 
\hline \hline
 & 610 MHz & 1280 MHz \\
\hline
Date of Obs. & 8 August 2014 & 14 August 2014 \\
Flux Calibrators & 3C286, 3C48 & 3C286, 3C48 \\
Phase Calibrators & 1714-252 & 1714-252 \\
On-source integration time  & $\sim$4~h & $\sim$4~h \\
Synth. beam & 10.69\arcsec$\times$6.16\arcsec & 4.41\arcsec$\times$2.24\arcsec \\
Position angle. (deg) & 7.04 & 6.37 \\
{\it rms} noise (mJy/beam) & 0.41 & 0.07 \\
Int. Flux Density (Jy) & $14.1 \pm 1.4$  & $12.6 \pm 1.3$   \\
{\small (integrated within $3\sigma$ level)} & &  \\
\hline
\end{tabular}
\end{center}
\end{table}

\subsection{Other available data}\label{data_archive}
\subsubsection{Near-infrared data from 2MASS} \label{data_2mass}
NIR ($\rm JHK_s$) data for point sources around our region of interest has been obtained from the Two Micron All Sky Survey (2MASS) Point Source Catalog (PSC). Resolution of 2MASS images is $\sim$5.0 arcsec. We select those sources for which the ``read-flag'' values are 1 - 3 to ensure a sample with reliable photometry. This data is used to study the stellar population associated with IRAS 17149$-$3916.

\subsubsection{Mid-infrared data from Spitzer} \label{data_spitzer}

The MIR images towards IRAS 17149$-$3916 are obtained from the archives of the Galactic Legacy Infrared Midplane Survey Extraordinaire (GLIMPSE) survey of the {\it Spitzer} Space Telescope. We retrieve images towards the region in the four Infrared Array Camera (IRAC; \citealt{2004ApJS..154...10F}) bands (3.6, 4.5, 5.8, 8.0\,$\rm \mu m$). These images have an angular resolution of $\lesssim 2$ arcsec with a pixel size of $\sim 0.6$ arcsec. We utilize these images in our study to present the morphology of the MIR emission associated with the region.

\subsubsection{Far-infrared data from Herschel} \label{data_herschel}
The far-infrared (FIR) data used in this paper have been obtained from the {\it Herschel} Space Observatory archives. Level 2.5 processed 70 - 500\,$\rm \mu m$ images from Spectral and Photometric Imaging Receiver (SPIRE; \citealt{2010A&A...518L...3G}) and JScanam images from the Photodetector Array Camera and Spectrometer (PACS; \citealt{2010A&A...518L...2P}), that were observed as part of the {\it Herschel} infrared Galactic plane Survey (Hi-GAL; \citealt{2008A&A...481..345M}), were retrieved. We use the FIR data to examine cold dust emission and investigate the cold dust clumps in the regions.

\subsubsection{Atacama Large Millimeter Array archival data} \label{data_alma}
We make use of the 1.4\,mm (Band~6) continuum maps obtained from the archives of {\it Atacama Large Millimeter Array (ALMA)} to identify the compact dust cores associated with IRAS 17149$-$3916. These observations were made in 2017 (PI: A.Sanchez-Monge \#2016.1.00191.S) using the extended 12m-Array configuration towards four pointings, S61, S62, S63, and S64. Each of these pointings sample different regions of the IRAS 17149$-$3916 complex. The retrieved maps have an angular resolution of 1.4\,arcsec $\times$ 0.9\,arcsec and a pixel scale of 0.16\,arcsec.

\subsubsection{Molecular line data from MALT90 survey}

The Millimeter Astronomy Legacy Team 90 GHz survey (MALT90; \citealt{{2011ApJS..197...25F},{2013PASA...30...57J}}) was carried out using the Australia Telescope National Facility (ATNF) Mopra 22-m telescope with an aim to characterize molecular clumps associated with {\hiirs}. The survey dataset contains molecular line maps of more than 2000 dense cores lying in the plane of the Galaxy, the corresponding sources of which were taken from the ATLASGAL 870\,$\rm \mu m$ continuum survey. The MALT90 survey covers 16 molecular line transitions lying near 90 GHz with a spectral resolution of $0.11$ km s$^{-1}$ and an angular resolution of 38\,arcsec. In this study we use the optically thin $\rm N_2 H^+$ line spectra to carry out the virial analysis of the detected dust clumps associated with this complex.

\section{Results}
\label{results}
\subsection{Ionized gas} \label{ionized}
\begin{figure}
\centering
\includegraphics[scale=0.3]{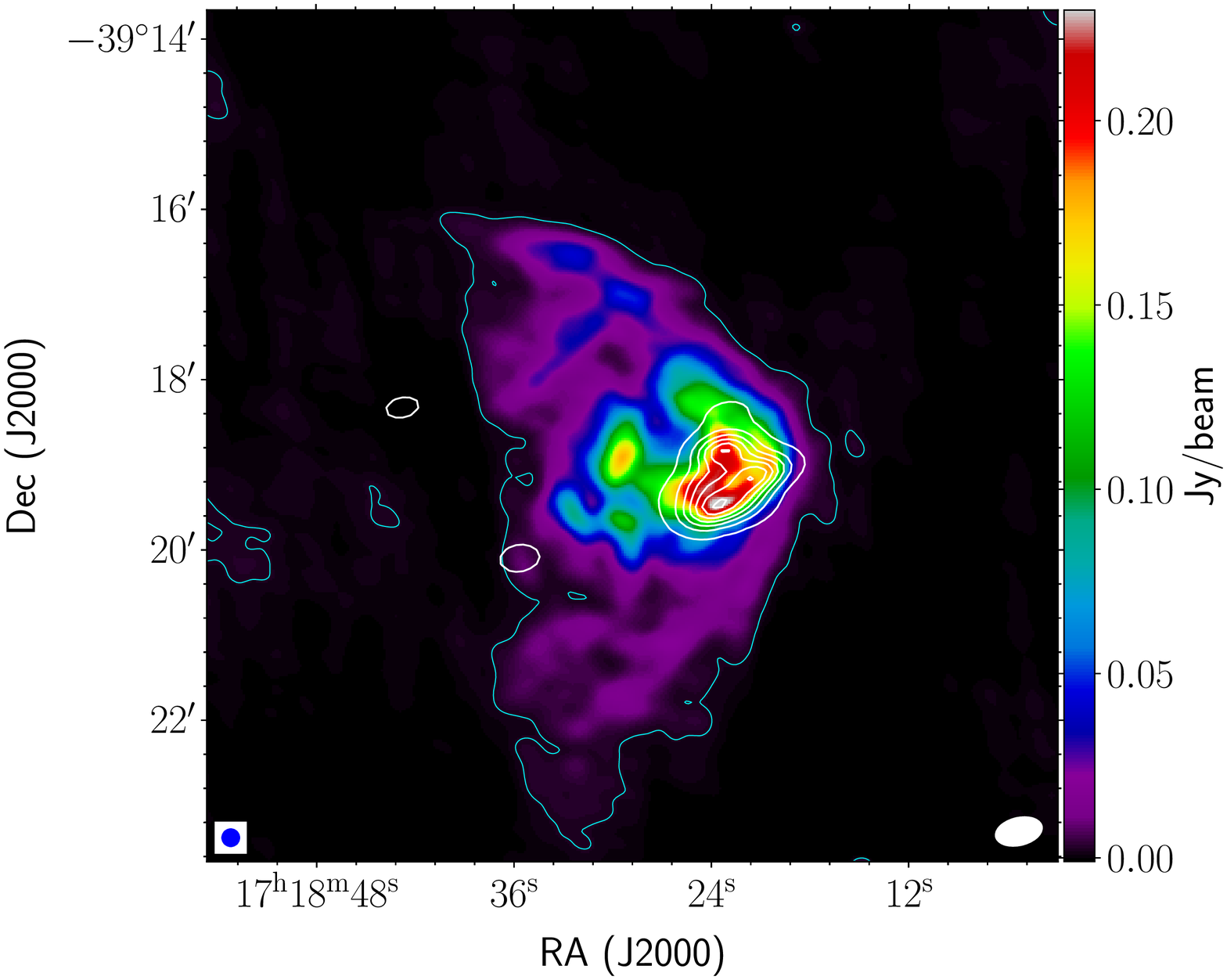}
\includegraphics[scale=0.3]{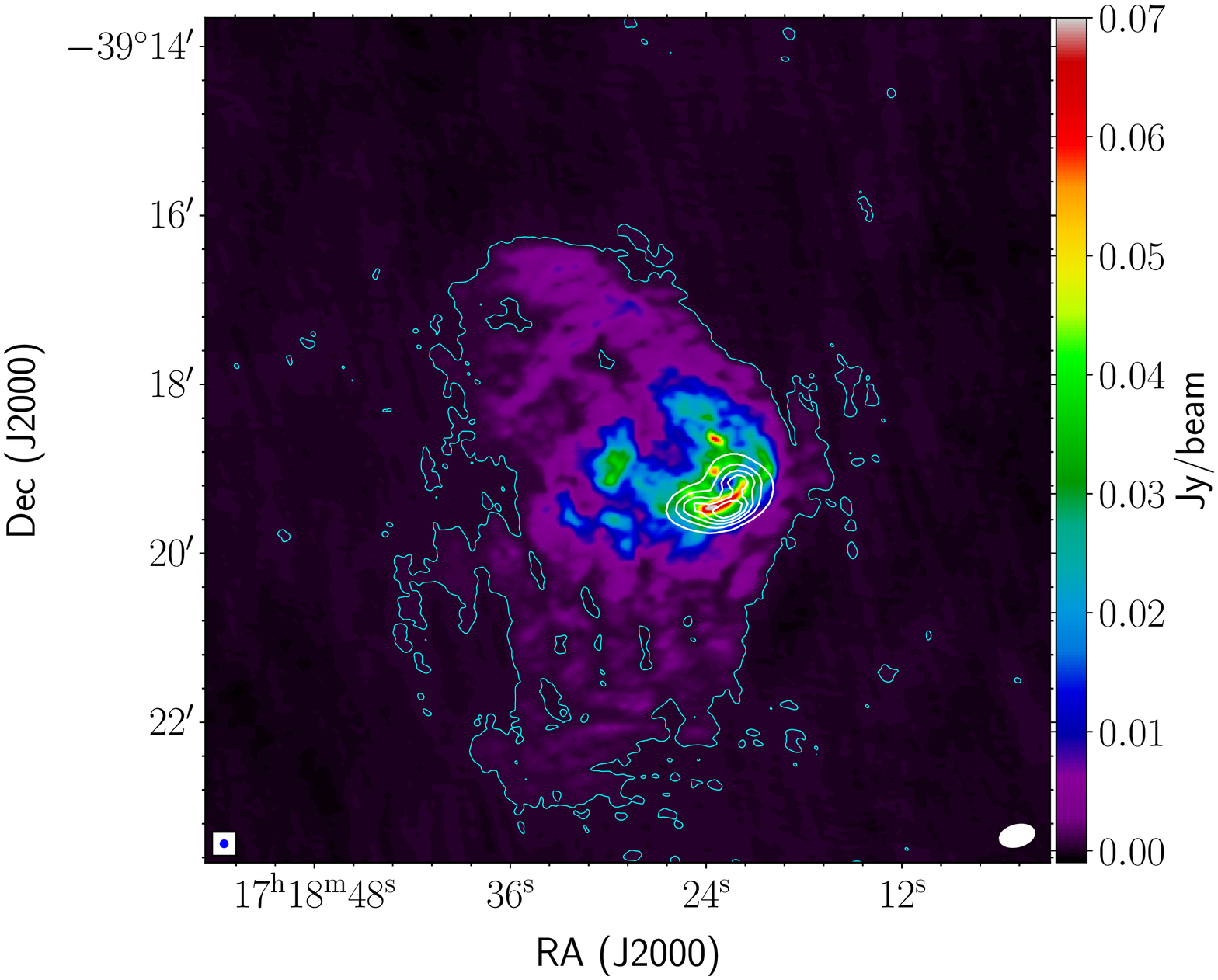}
\caption{{\it Top}: 610\,MHz GMRT map of the region associated with IRAS 17149$-$3916 overlaid with the 3$\sigma$ ($\sigma = 0.6\,\rm mJy\,beam^{-1}$) level contour in light blue. White contours correspond to the 18\,GHz emission mapped with ATCA \citep{2013A&A...550A..21S}, with contour levels starting from 3$\sigma$ and increasing in steps of 8$\sigma$ ($\sigma = 1.2\times 10^{-2}\,\rm Jy\,beam^{-1}$). {\it Bottom:} Same as the top panel but for 1280\,MHz GMRT ($\sigma = 0.1\,\rm mJy\,beam^{-1}$) and 22.8\,GHz ATCA map with the contour levels starting from 3$\sigma$ and increasing in steps of 5$\sigma$ ($\sigma = 1.2\times 10^{-2}\,\rm Jy\,beam^{-1}$). The GMRT maps presented here have been convolved to a resolution of 12\,arcsec for 610~MHz and 5\,arcsec for 1280~MHz. The beams sizes of the GMRT and ATCA maps are shown in the lower left and right hand corner, respectively.}
\label{radioim}
\end{figure}
We present the first low-frequency radio maps of the region associated with IRAS 17149$-$3916 obtained using the GMRT. The continuum emission mapped at 610 and 1280~MHz are shown in Fig.~\ref{radioim}. The ionized gas reveals an interesting, large-extent cometary morphology where the head lies in the west direction with a fan-like diffuse tail opening in the east. The tail has a north-south extension of $\sim$ 6 arcmin. The bright radio emission near the head displays a `horse shoe' shaped structure that opens towards the north-east and mostly traces the south-western portion of the dust ring structure presented in Fig.~\ref{fig_intro}. This is enveloped within the extended and faint, diffuse emission. In addition, there are several discrete emission peaks seen at both frequencies. The {\it rms} noise and the integrated flux density values estimated are listed in Table \ref{radio_tab}. For the latter, the flux density is integrated within the respective 3$\sigma$ contours. The quoted errors are estimated following the method discussed in \citet{2018A&A...612A..36D}.

Also included in the figure are contours showing the high-frequency ATCA observations at 18 and 22.8~GHz, from \citet{2013A&A...550A..21S}. These snapshot ($\sim$ 10~mins) ATCA maps sample only the brightest region towards the head and the emission at 18~GHz is seen to be more extended. The GMRT and ATCA maps reveal the presence of several distinct peaks which are likely to be externally ionized density enhanced regions thus suggesting a clumpy medium. 
The fact that some of these peaks could also be internally ionized by newly formed massive stars cannot be ruled out. Hence, further detailed study is required to understand the nature of these radio peaks. 
A careful search of the SIMBAD/NED database rules out any possible association with background/foreground radio sources in the line of sight. Comparing with Fig.~\ref{fig_intro}, the ionized emission traced in the $\rm H{\alpha}$ image agrees well with the GMRT maps.  \citet{2006AJ....131..951R} and \citet{2014MNRAS.437..606T} present the ionized emission mapped in the Br$\gamma$ line which is localized to the central part and mostly correlates with the bright emission seen in the GMRT maps. 

Assuming the radio emission at 1280~MHz to be optically thin and emanating
from a homogeneous, isothermal medium, we derive several physical parameters of the detected {\hiir} using the following expressions from \citet{2016A&A...588A.143S},\\

{\it Lyman continuum photon flux ($N_{\rm Ly}$):}

\begin{equation}
\begin{split}
\left( \frac{N_{\rm Ly}}{\rm s^{-1}}\right) = 4.771 \times 10^{42} \left(\frac{F_\nu}{\rm Jy}\right) \left( \frac{T_{\rm e}}{\rm K}\right)^{-0.45} \\
\times \left( \frac{\nu}{\rm GHz}\right)^{0.1} \left( \frac{D}{\rm pc}\right) ^{2}
\end{split}
\label{Lyman_flux}
\end{equation}

{\it Electron number density ($n_{\rm e}$):}
\begin{equation}
\begin{split}
\left ( \frac{n_{\rm e}}{\rm cm^{-3}}\right)  = 2.576 \times 10^6 \left(\frac{F_\nu}{\rm Jy}\right)^{0.5} \left( \frac{T_{\rm e}}{\rm K}\right)^{0.175} \left( \frac{\nu}{\rm GHz}\right)^{0.05} \\
\times   \left( \frac{\theta_{\rm source}}{\rm arcsec}\right)^{-1.5} \left( \frac{D}{\rm pc}\right) ^{-0.5}
\end{split}
\label{e_no_density}
\end{equation}

{\it Emission measure (EM):}
\begin{equation}
\begin{split}
\left( \frac{\rm EM}{\rm pc\ cm^{-6}}\right) = 3.217 \times 10^7  \left(\frac{F_\nu}{\rm Jy}\right) \left( \frac{T_{\rm e}}{\rm K}\right)^{0.35} \\
\times \left( \frac{\nu}{\rm GHz}\right)^{0.1} \left( \frac{\theta_{\rm source}}{\rm arcsec}\right)^{-2}
\end{split}
\label{emission_measure}
\end{equation}
where, $F_{\nu}$ is the integrated flux density of the ionized region, $T_{\rm e}$ is the electron temperature, $\nu$ is the frequency, $\theta_{\rm source}$ is the angular diameter of the {\hiir} and D is the distance to this region. $T_{\rm e}$ is taken to be 5000~K from the radio recombination line estimates by \citet{1987A&A...171..261C}. Approximating the emission region to an ellipse, the angular source size ($\theta_{\rm source}$) is taken to be the geometric mean of the axes of the ellipse and is estimated to be 6.25~arcmin (3.6~pc).  The derived physical parameters are listed in Table \ref{radio-physical-param}. 

\begin{table}
\caption{Derived physical parameters of the {\hiir} associated with IRAS 17149$-$3916.}
\begin{center}
\begin{tabular}{ccccc}
\hline
Size & log $N_{\rm Ly}$ & EM            & $n_{\rm e}$    & Spectral Type \\
(pc) &              & (cm$^{-6}$pc) & (cm$^{-3}$) &               \\
\hline      

3.6  & 48.73        & 5.8$\times$10$^{5}$ & 1.3$\times$10$^{2}$ &  O6.5V -- O7V \\ \hline
\end{tabular}
\label{radio-physical-param}
\end{center}
\end{table}
 
If we assume a single star to be ionizing the \hiir\ and compare the Lyman-continuum photon flux obtained from the 1280~MHz map with the parameters of O-type stars presented in \citet[Table 1;][]{2005A&A...436.1049M}, we estimate its spectral type to be O6.5V -- O7V. 
This can be considered as a lower limit as the emission at 1280~MHz could be optically thick as well. In addition, one needs to account for dust absorption of Lyman continuum photons, which can be significant as shown by many studies \citep[e.g.][]{2011A&A...525A.132P}. 
The estimated spectral type suggests a mass range of $\sim 20 - 40~ M_{\odot}$ for the ionizing star \citep{2005A&A...436.1049M}. 

To decipher the nature of the ionized emission, we determine the spectral index, $\alpha$ which is defined as $F_{\nu} \propto \nu^{\alpha}$. The flux density, $F_{\nu}$, is calculated from the GMRT radio maps. For this, we generate two new radio maps at 610 and 1280~MHz by setting the $\it uv$ range to a common value of $0.14 - 39.7 $~k$\lambda$. This ensures similar spatial scales being probed at both frequencies. Further, the beam size for both the maps is set to $\rm 12~arcsec \times 12~arcsec$. $F_{\nu}$ is obtained by integrating within the area defined by the 3$\sigma$ contour of the new 610~MHz map. 
The integrated flux density values are estimated to be $\rm 13.7 \pm 1.3 \,Jy\,$, $\rm 12.1 \pm 1.2 \,Jy\,$  at 610 and 1280~MHz, respectively. 
These yield a spectral index of $-0.17 \pm 0.19$. Similar values are obtained for the central, bright radio emission as well. Within the quoted uncertainties, the average spectral index is fairly consistent with optically thin, free-free emission as expected from {\hiirs} which are usually dominated by thermal emission. Spectral index estimate of $-0.1$, consistent with optically thin thermal emission, is also obtained by combining the GMRT flux density values with the available single dish measurements at 2.65~GHz \citep{1969AuJPA..11...27B} and 4.85~GHz \citep{1994ApJS...91..111W}.

\subsection{The dust environment}
\label{mir-dust}
\begin{figure*}
\centering
\includegraphics[scale=0.4]{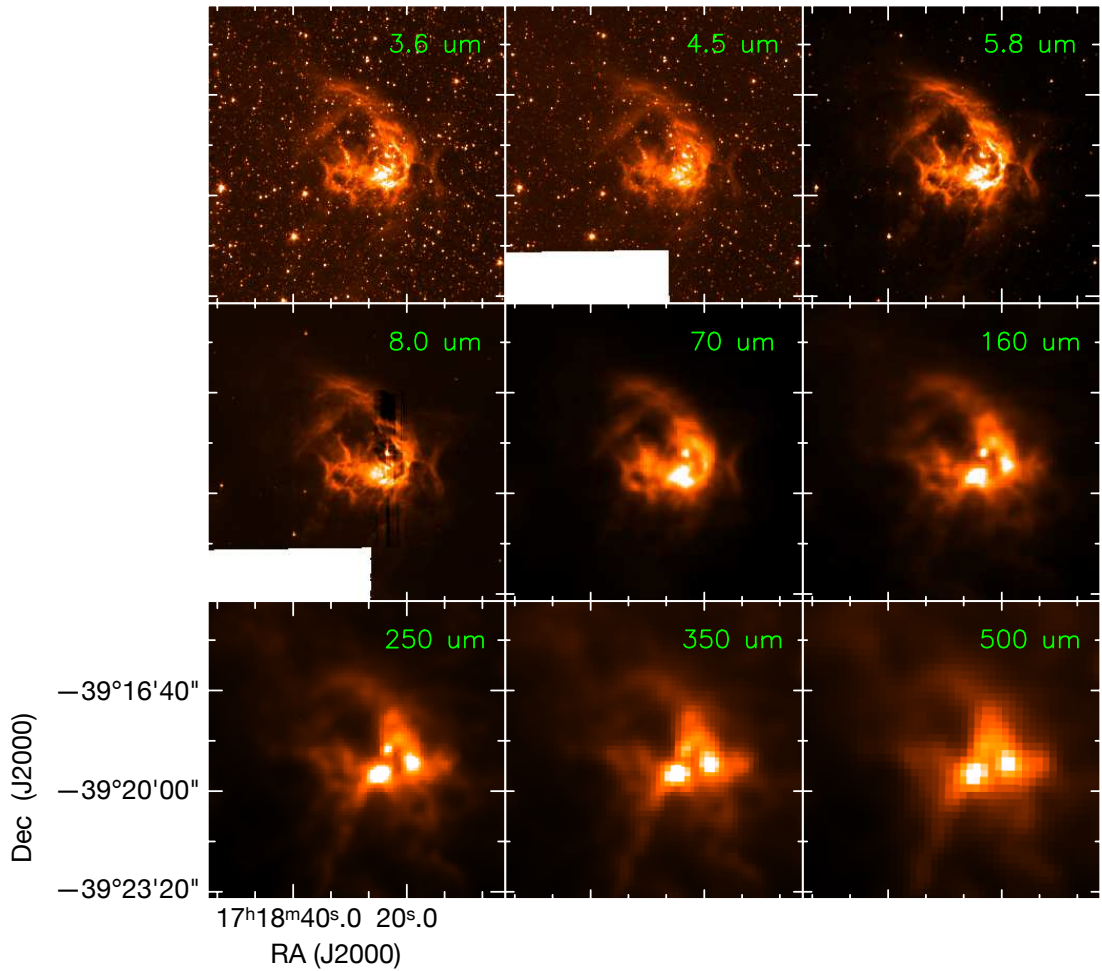}
\caption{Dust emission associated with IRAS 17149$-$3916 from the {\it Spitzer}-GLIMPSE and the {\it Herschel}-Hi-GAL surveys.} 
\label{iremission}
\end{figure*}
The warm and the cold dust emission associated with IRAS 17149$-$3916 unravels interesting morphological features like a bubble, pillars, filaments, arcs, and clumps, that strongly suggest this to be a very active star forming complex where the profound radiative and mechanical feedback of massive stars on the surrounding ISM is clearly observed. Fig.~\ref{iremission} compiles the MIR and FIR emission towards IRAS 17149$-$3916 from the GLIMPSE and Hi-GAL surveys. Apart from the stellar population probed at the shorter wavelengths, the diffuse emission seen in the IRAC-GLIMPSE images would be dominated by the emissions from polycyclic aromatic hydrocarbons (PAHs) excited by the UV photons in the photodissociation regions \citep{2012A&A...542A..10A,2012ApJ...760..149P}. Close to the hot stars, there would also be significant contribution from thermally emitting warm dust that is heated by stellar radiation \citep{2008ApJ...681.1341W}.  In {\hiirs}, emission from dust heated by trapped Ly$\alpha$ photons \citep{1991MNRAS.251..584H}, would also be present in these IRAC bands. In the wavelength regime of the 21~$\rm \mu m$ MSX, the emission is either associated with stochastically heated Very Small Grains (VSGs) or thermal emission from hot big grains (BGs). As we go to the FIR Hi-GAL maps, cold dust emission dominates and shows up as distinct clumps and filamentary structures where, emission in the 70~$\rm \mu m$ band is dominated by the VSGs and the longer wavelength bands like 250~$\rm \mu m$ band trace emissions from the BGs \citep{2012ApJ...760..149P}.

\subsubsection*{Dust temperature and column density maps}
To understand the nature of cold dust emission, we generate the dust temperature and the molecular hydrogen column density maps following the procedure detailed in \citet{2018A&A...612A..36D} and \citet{2019MNRAS.485.1775I} and briefly stated here. A pixel-by-pixel modified blackbody modelling to the observed spectral energy distribution is carried out. As discussed in these two papers, the 70\,$\rm \mu m$ data is not used because there would be appreciable contribution from the warm dust component hence rendering a single modified blackbody fit inaccurate. Thus, we have the FIR emission at 160, 250, 350, and 500\,$\rm \mu m$ mostly on the Rayleigh-Jeans part to constrain the model given by
\begin{equation}
F_{\nu}-I_{\rm bg} = B_{\nu}(\nu,T_{\rm d})~\Omega~(1-{\rm e}^{-\tau_{\nu}}) 
\label{MBB-Eqn}
\end{equation}
where,  $F_{\nu}$ is the observed flux density, $I_{\rm bg}$ is the background flux density,  $B_{\nu}(\nu,T_{\rm d})$ is the Planck function at the dust temperature $\rm T_{\rm d}$, $\Omega$ is the solid angle subtended by a pixel (all maps are convolved to a common resolution of 35.7~arcsec and regridded to a common pixel size of  $\rm 14~arcsec\times14~arcsec$). The background flux is estimated from a nearby region relatively free of clumpy and bright emission.
The optical depth $\tau_{\nu}$ in Eqn. \ref{MBB-Eqn} can be expressed as
\begin{equation}
\tau_{\nu} = \mu_{\rm H_2}~ N({\rm H_2})~ m_{\rm H}~ \kappa_{\nu}
\end{equation}
where, $\mu_{\rm H_2}$ is the mean molecular weight which is taken as 2.8 \citep{2008A&A...487..993K}, $N({\rm H_2})$ is the column density, $m_{\rm H}$ is the mass of hydrogen atom and $\kappa_{\nu}$ ($\rm cm^2\,g^{-1}$) is the dust opacity which is given as \citep{1983QJRAS..24..267H}
\begin{equation}
\kappa_{\nu} = 0.1\left ( \frac{\nu}{1200~{\rm GHz}} \right )^{\beta} 
\label{kappa}
\end{equation}
Here, $\beta$ denotes the dust emissivity spectral index and a typical value of 2 estimated in several star regions is assumed. 
In fitting the modified blackbody to the observed flux densities, $N({\rm H_2})$ and $T_{\rm d}$ are kept as free parameters. 
\begin{figure*}
\centering
\includegraphics[scale=0.32]{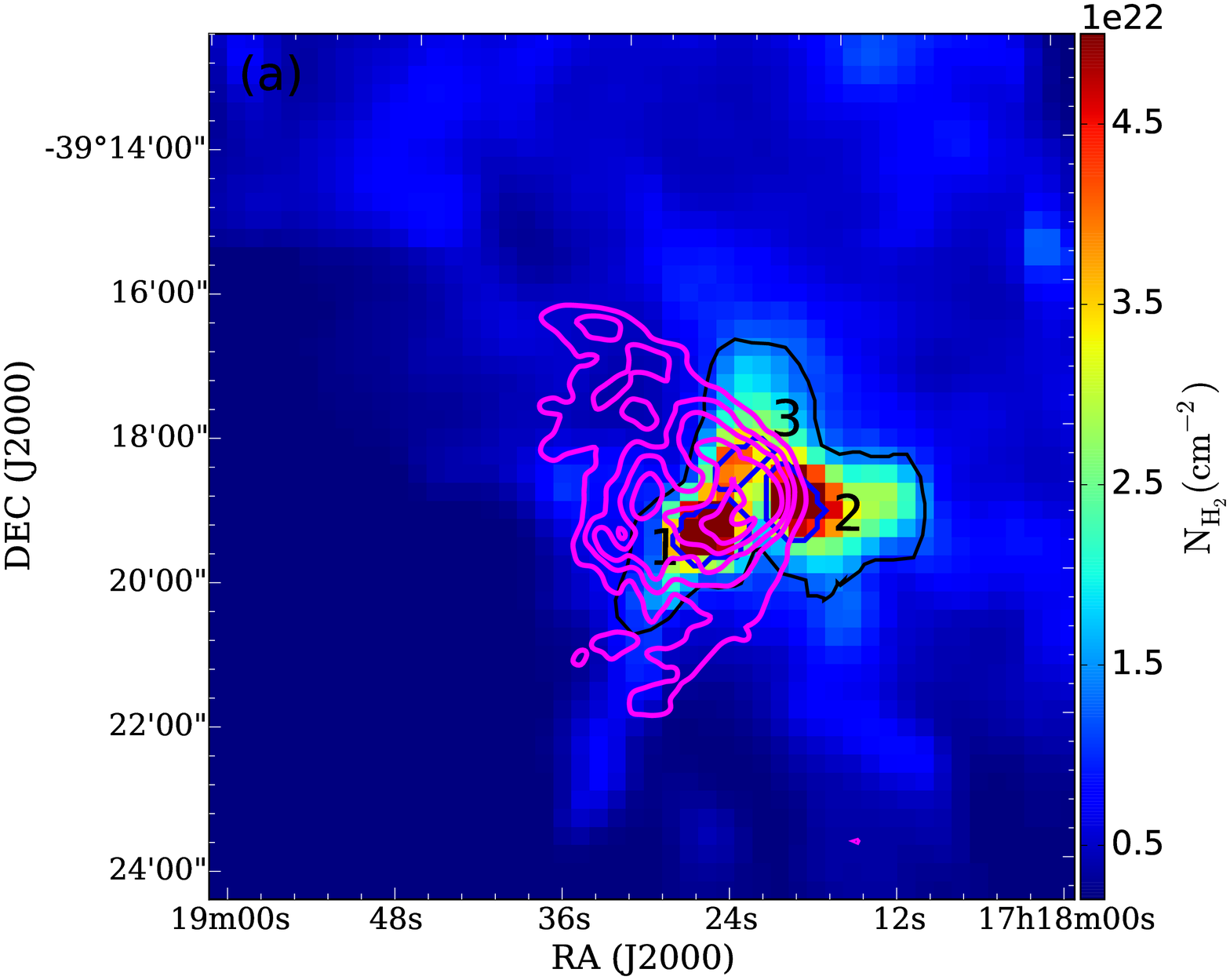}
\includegraphics[scale=0.32]{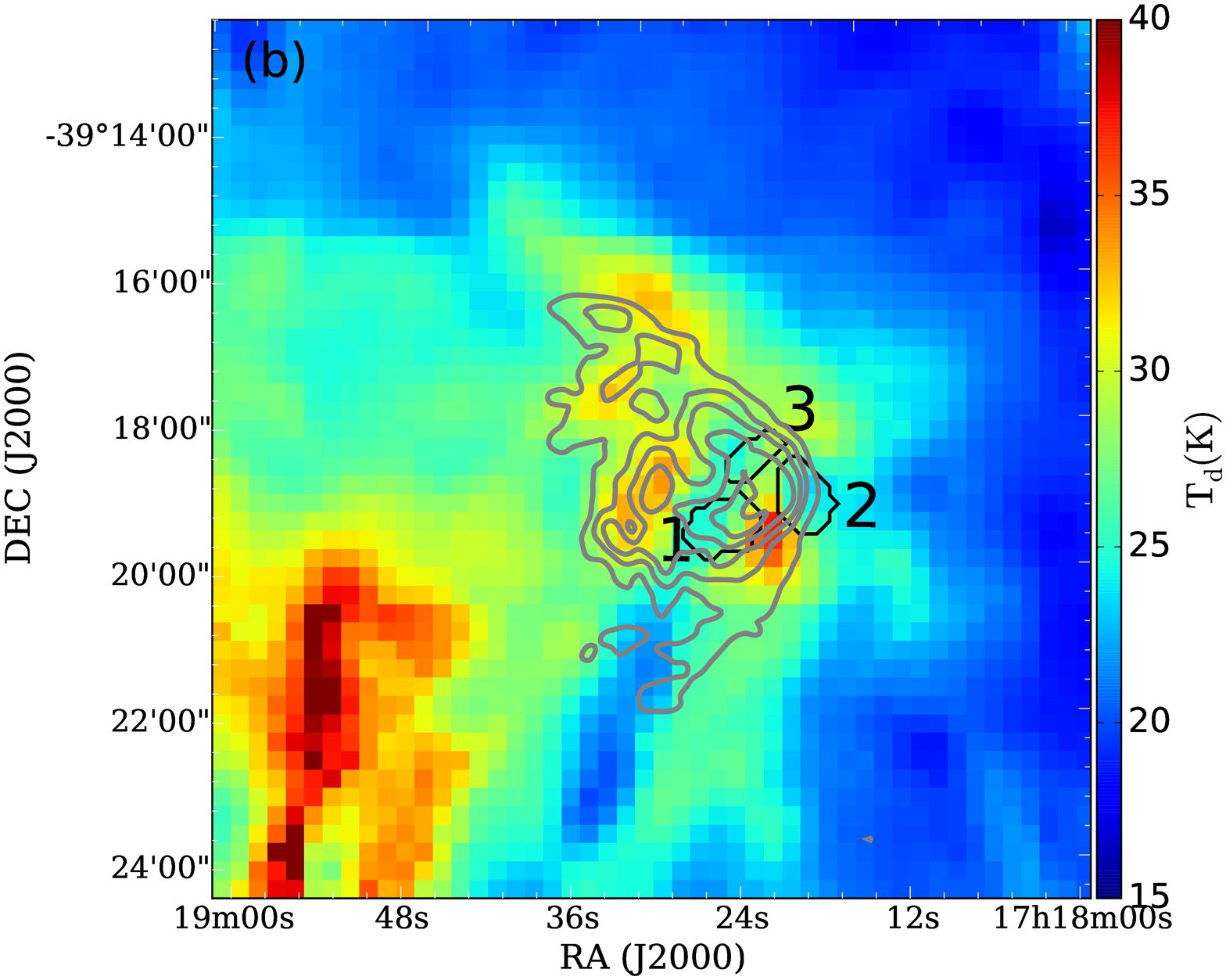}
\includegraphics[scale=0.32]{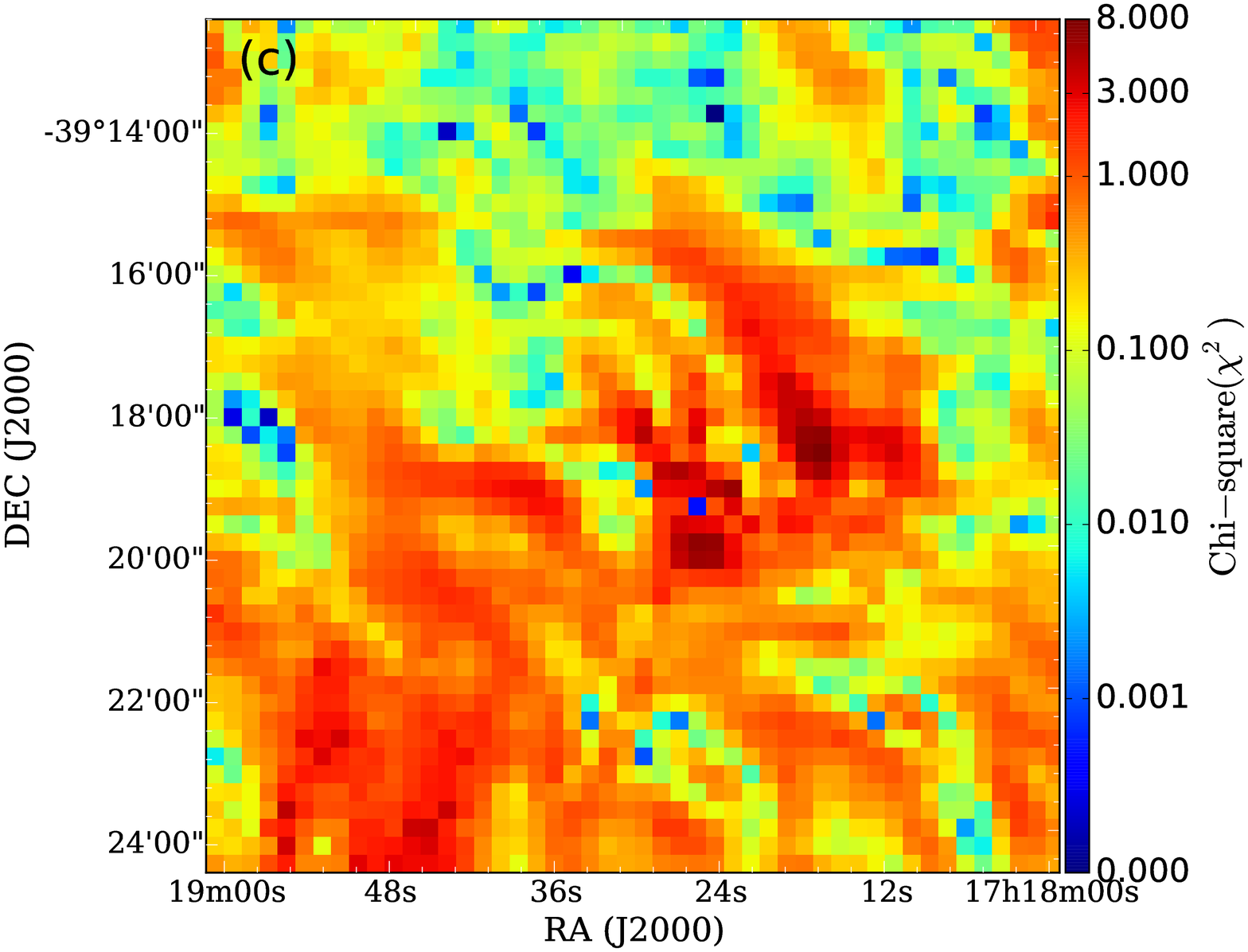}
\caption{Column density (a), dust temperature (b), and chi-square ($\chi^{2}$) (c) maps of the region associated with the \hii\ region. The 610~MHz ionized emission overlaid as magenta and gray contours on column density and dust temperature maps, respectively. The contour levels are 0.01, 0.03, 0.06, 0.12, and 0.2~Jy/beam. The retrieved clump apertures are shown on the column density (in blue) and dust temperature (in black) maps following the nomenclature as discussed in the text. The black contour in (a) shows the area integrated for estimating $n_0$ (refer Section \ref{clumps-CC}.)}
\label{cdtchi}
\end{figure*}
The column density and dust temperature maps generated are shown in Fig.~\ref{cdtchi}. The goodness of the fits for each pixel can be seen in the $\chi^{2}$ map where the maximum $\chi^{2}$ value is seen to be $\sim 8$.
The column density map presents a triangular morphology with three distinct, bright and dense regions.
A network of broad filaments are also seen in the map. The dust temperature map is relatively patchy with regions of higher temperature within the radio nebula. A region with warm temperature is seen to be located towards the south-east of IRAS 17149$-$3916, the signature of which can be seen in the {\it Herschel} maps shown in Fig.~\ref{iremission}. The western side of the {\hiir} shows comparatively cold temperatures. Furthermore, the filamentary features seen in the column density map are mostly revealed as distinct low temperature lanes. 

\subsubsection*{Dust clumps and cores}\label{dust_clump}
\begin{figure*}
\centering
\includegraphics[scale=0.5]{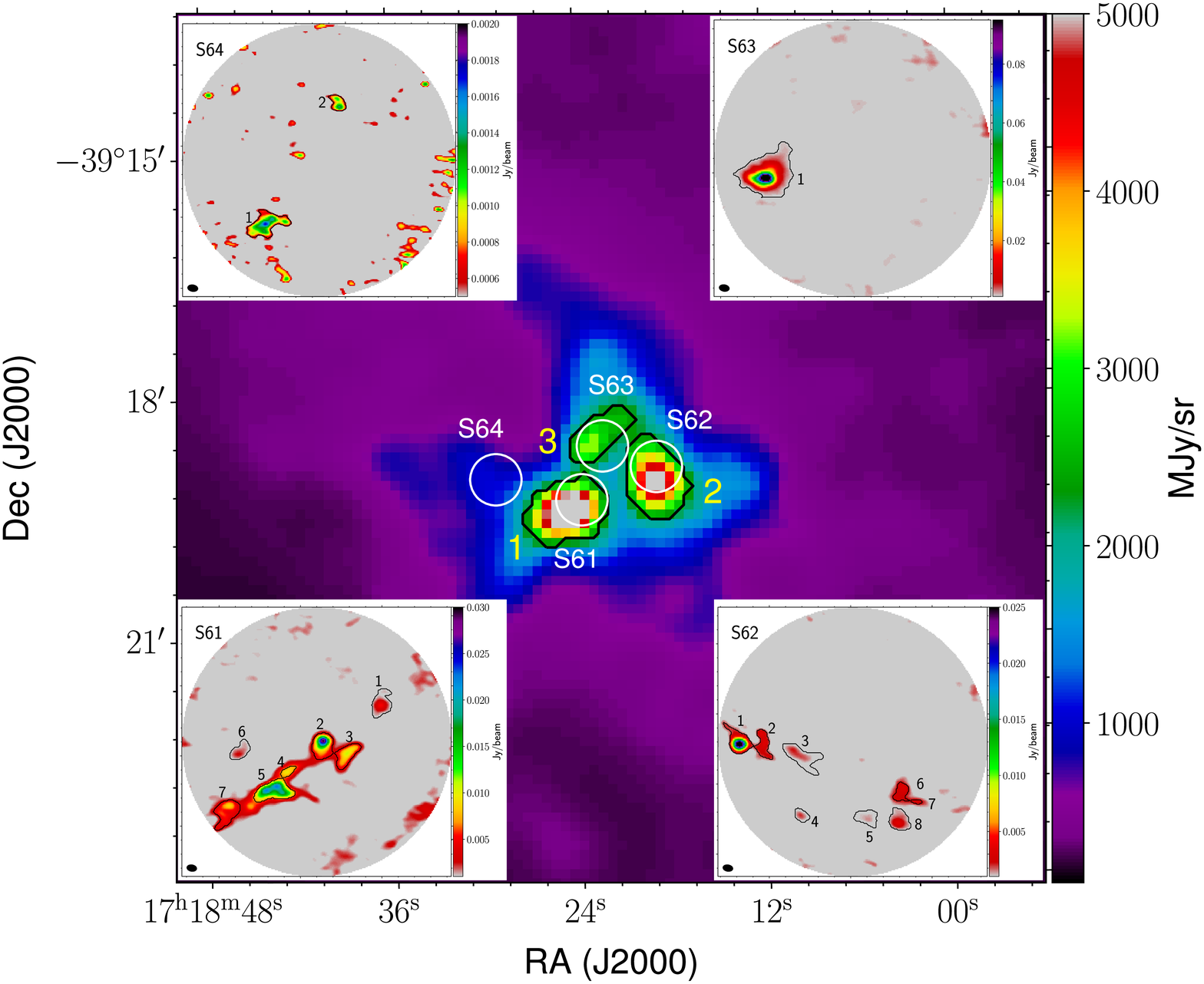}
\caption{ The central panel is the \textit{Herschel} 350\,$\rm \mu m$ map overlaid with the {\it Dendrogram} retrieved clump apertures in black. White circles represent four pointings of the {\it ALMA} observations, S61, S62, S63, and S64. The 1.4\,mm {\it ALMA} map towards each pointing is placed at the corners of the 350\,$\rm \mu m$ image. The apertures of the dense cores extracted using the {\it Dendrogram} algorithm are overlaid and labelled on these maps. The beam sizes of the 1.4\,mm maps are given towards the lower left hand corner of each image.}
\label{dustclumps}
\end{figure*}

The FIR and the column density maps show the presence of dust clumps. These clumps are identified using the \textit{Herschel} 350\,$\rm \mu m$ map and the {\it Dendrogram}\footnote{\url{https://dendrograms.readthedocs.io/en/stable/}} algorithm. Using this algorithm, we identify the smallest structures, called the `leaves', in the 350\,$\rm \mu m$ map, which in this case are the cold dust clumps. The key input parameters for the identification of the clumps are (1) {\it min\_value} = $3\sigma$ and (2) {\it min\_delta} = $\sigma$, where $\rm \sigma (= 191.2\,MJy\,sr^{-1})$ is the {\it rms} level of the 350\,$\rm \mu m$ map. An additional parameter, {\it min\_pix = N}, is also used, which is the minimum number of pixels required for a `leaf' to be considered an independent entity. To ensure that the clumps are resolved, the value of {\it N} is chosen to be 7\,pixels, the beam area of the 350\,$\rm \mu m$ map. Setting these parameters, we extract three cold dust clumps. The central panel of Fig.~\ref{dustclumps} shows the 350\,$\rm \mu m$ map overlaid with the retrieved apertures of the three detected clumps labelled 1, 2, and 3. The physical parameters of the detected clumps are listed in Table \ref{clump-param}. These are derived from the 350\,$\rm \mu m$, column density and dust temperature maps. The peak positions are determined from the 350\,$\rm \mu m$ map. The clump radii, $r=(A/\pi)^{0.5}$, where $A$ is the enclosed area within the retrieved clump apertures. 
\begin{table*}
\caption{Physical parameters of the detected dust clumps associated with IRAS 17149$-$3916}
\begin{center}
\centering 
\scalebox{0.85}{
\begin{tabular}{ccccccccccc}
\hline
Clump & \multicolumn{2}{c}{Peak position} & Mean $T_{\rm d}$ & $\Sigma N({\rm H_2})$       & Radius & Mass          & Mean $N({\rm H_2})$         & No. density $(n_{\textup{H}_{2}})$ & \multicolumn{1}{l}{$M_{\rm vir}$} & \multicolumn{1}{l}{$\alpha_{\rm vir}$} \\
      & RA (J2000)      & DEC (J2000)     & (K)              & $(\times 10^{23}$ cm$^{-2})$ & (pc)   & ($M_{\odot}$) & $(\times 10^{22}$ cm$^{-2}$) & $(\times 10^{4}$ cm$^{-3}$)        & ($M_{\odot}$)                     & \multicolumn{1}{l}{}                   \\ \hline
1     & 17 18 24.32     & -39 19 28.04    & 27.8             & 6.1                          & 0.3    & 250           & 5.5                          & 5.0                                & 452                               & 1.75                                   \\
2     & 17 18 18.77     & -39 19 04.54    & 25.0             & 7.1                          & 0.3    & 292           & 5.4                          & 5.1                                & 600                               & 2.15                                   \\
3     & 17 18 22.57     & -39 18 36.44    & 27.3             & 2.8                          & 0.2    & 117           & 3.6                          & 4.6                                & 450                               & 4.08                                   \\ \hline
\end{tabular}}
\label{clump-param}
\end{center}
\end{table*}
To estimate the masses of the detected clumps, we utilize the column density map and use the following expression
\begin{equation}
M_{\rm clump} =\mu_{\rm H_2}~ \Sigma N({\rm H_2})~A_{\rm pixel}~m_{\rm H}
\label{mass_eqn}
\end{equation}
where, $\mu_{\rm H_2}$ is the mean molecular weight taken as 2.8, $\Sigma N({\rm H_2})$ is the integrated column density over the clump area, $A_{\rm pixel}$ is the pixel area in $\rm cm^2$ and $m_{\rm H}$ is the mass of hydrogen atom. The number density is determined using the expression $n_{\rm H_2} = 3\,N({\rm H_2})/4r$. The peak positions of the Clumps 1, 2, and 3 agree fairly well with Clumps III, I, and II, respectively, detected by \citet{2014MNRAS.437..606T} using {\it Herschel} maps. Comparing the masses presented in Table \ref{clump-param}, the estimates given in \citet{2014MNRAS.438.2716T} are higher by a factor of 8, 1.5, and 3 for the Clumps 1, 2, and 3, respectively. 

{\it ALMA} continuum data enables investigation of the detected clumps at high resolution.  In Fig.~\ref{dustclumps}, we show {\it ALMA} dust continuum maps at 1.4\,mm towards the four pointings marked and labelled as S61, S62, S63, and S64 where the first three pointings lie mostly within three clumps and S64 lies outside towards the east. Using the same {\it Dendrogram} algorithm, several cores are identified. The key input parameters to the {\it Dendrogram} algorithm are, {\it min\_value} = $3\sigma$, {\it min\_delta} = $\sigma$, and {\it min\_pix = N}, where $\sigma$ is the {\it rms} level and $N(=60)$ is the beam area of the 1.4\,mm maps. In order to avoid detection of spurious cores, we only retain those with peak flux density greater than $5\sigma$. Applying these constraints, seven cores are identified towards S61 and S62 each, one towards S63 and two towards S64. 

To further study these dense cores, we estimate their physical parameters. Adopting the formalism described by \citet{2018ApJ...853..160C} and assuming the emission at 1.4\,mm to be optically thin, the masses are estimated using the following equation

\begin{eqnarray}
  M & = &
  \displaystyle 0.0417 \, M_{\odot}
  \left( {\textrm e}^{0.533 (\lambda / {1.3\, \textrm {mm}})^{-1}
      (T / {20\, \textrm {K}})^{-1}} - 1 \right) \left( \frac{F_{\nu}}{\textrm {mJy}} \right) \nonumber \\
  & & \displaystyle
 \times \left( \frac{\kappa_{\nu}}{0.00638\,\textrm{cm}^2\,\textrm{g}^{-1}} \right)^{-1}
  \left( \frac{d}{\textrm {kpc}} \right)^2
  \left( \frac{\lambda}{1.3\, \textrm {mm}} \right)^{3} 
  \label{core_mass}
\end{eqnarray}
Here, $F_\nu$ is the integrated flux density of each core, $d$ is the distance to the source and $\lambda$ is the wavelength. Opacity, $\kappa_\nu$ is estimated using equation~\ref{kappa} with the dust emissivity spectral index $\beta$ fixed at 2.0. For cores detected in S61, S62, and S63, mean dust temperatures of the respective clumps are taken. For S64, the mean dust temperature for the region covering the S64 pointing is used. The effective radius, $r=(A/\pi)^{0.5}$, of each core is also estimated where $A$ is area enclosed within each core aperture.
The identified cores with the retrieved apertures are shown in Fig.~\ref{dustclumps} and the estimated physical parameters are list in Table \ref{alma-cores}. The uncertainties related to the missing flux effect are not taken into account in deriving these parameters since it may not be significant given that the largest recoverable scales is quoted to be $\sim$10~arcsec for this ALMA dataset which is appreciably larger than the typical sizes of the detected cores. Barring the largest detected core which has an angular size of $\sim$7~arcsec, the average size of the cores is 3~arcsec.
\begin{table*}
\caption{Parameters of the detected dust cores extracted associated with IRAS 17149$-$3916}
\begin{center}
\centering 
\begin{tabular}{ccccccc}
\hline
    & Core & \multicolumn{2}{c}{Peak position}    &  Flux density  & Radius     & Mass \\
    &   & RA (J2000)      & DEC (J2000)       &  (mJy)    & (pc)      & ($M_\odot$)       \\
\hline
S61 & 1 & 17 18 23.03   & -39 19 12.95      & 9.7   & 0.01  & 1.7   \\
    & 2 & 17 18 23.75   & -39 19 18.47      & 62.2  & 0.02  & 10.7  \\
    & 3 & 17 18 23.45   & -39 19 19.53      & 32.8  & 0.02  & 5.7   \\
    & 4 & 17 18 24.17   & -39 19 22.35      & 11.2  & 0.01  & 1.9   \\
    & 5 & 17 18 24.34   & -39 19 25.09      & 106.7 & 0.02  & 18.4  \\
    & 6 & 17 18 24.76   & -39 19 19.50      & 4.5   & 0.01  & 0.8  \\
    & 7 & 17 18 24.93   & -39 19 28.12      & 31.8  & 0.02  & 5.5   \\
\hline 
S62 & 1 & 17 18 20.38   & -39 18 54.32      & 46.0  & 0.02  & 9.4  \\
    & 2 & 17 18 20.09   & -39 18 54.43      & 9.8   & 0.01  & 2.0  \\
    & 3 & 17 18 19.64   & -39 18 56.07      & 7.9   & 0.02  & 1.6  \\
    & 4 & 17 18 19.63   & -39 19 04.70      & 2.1   & 0.01  & 0.4  \\
    & 5 & 17 18 18.82   & -39 19 04.94      & 3.1   & 0.01  & 0.6   \\
    & 6 & 17 18 18.35   & -39 19 01.65      & 19.2  & 0.02  & 3.9  \\
    & 7 & 17 18 18.42   & -39 19 05.37      & 8.2   & 0.01  & 1.7   \\
\hline 
S63 & 1 & 17 18 23.48   & -39 18 40.70      & 357.7 & 0.03  & 62.3  \\
\hline 
S64 & 1 & 17 18 29.98   & -39 19 11.22      & 7.0   & 0.02  & 1.1  \\
    & 2 & 17 18 29.10   & -39 18 54.11      & 1.9   & 0.01  & 0.3  \\

\hline
\end{tabular}
\label{alma-cores}
\end{center}
\end{table*}

\subsection{Molecular line observation of identified clumps}
\begin{figure*}
\centering
\includegraphics[scale=0.2]{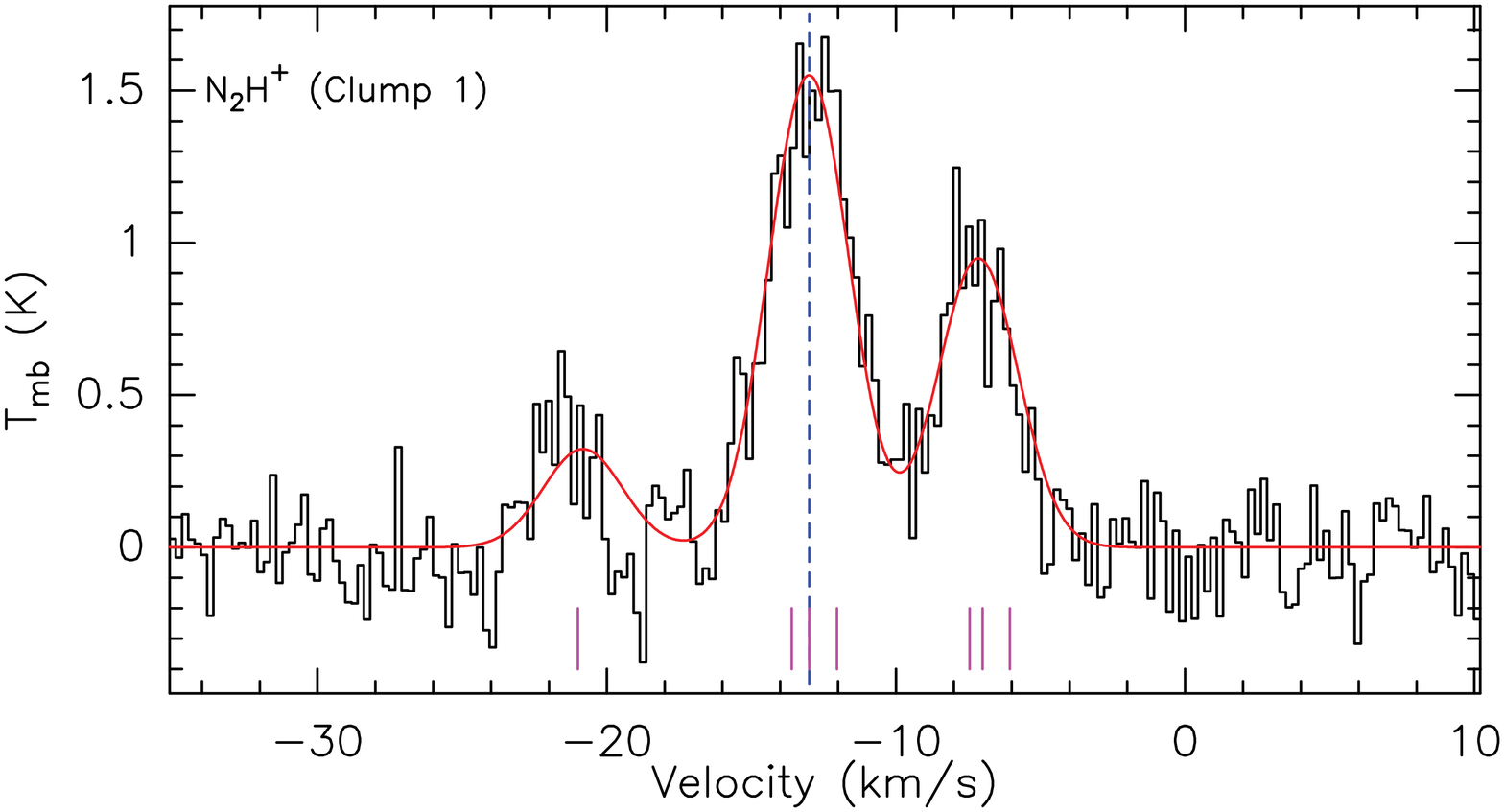}
\includegraphics[scale=0.2]{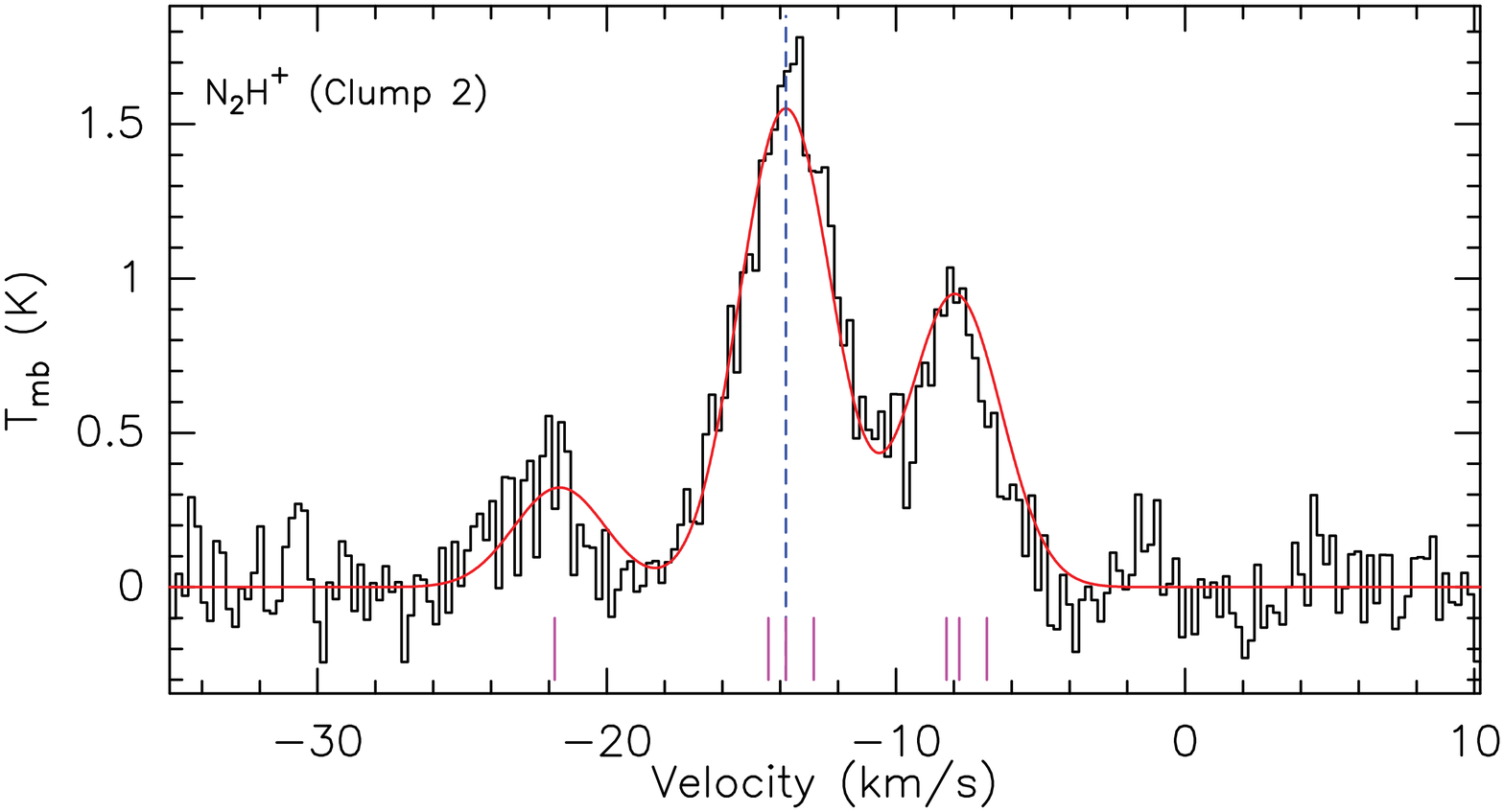}
\includegraphics[scale=0.2]{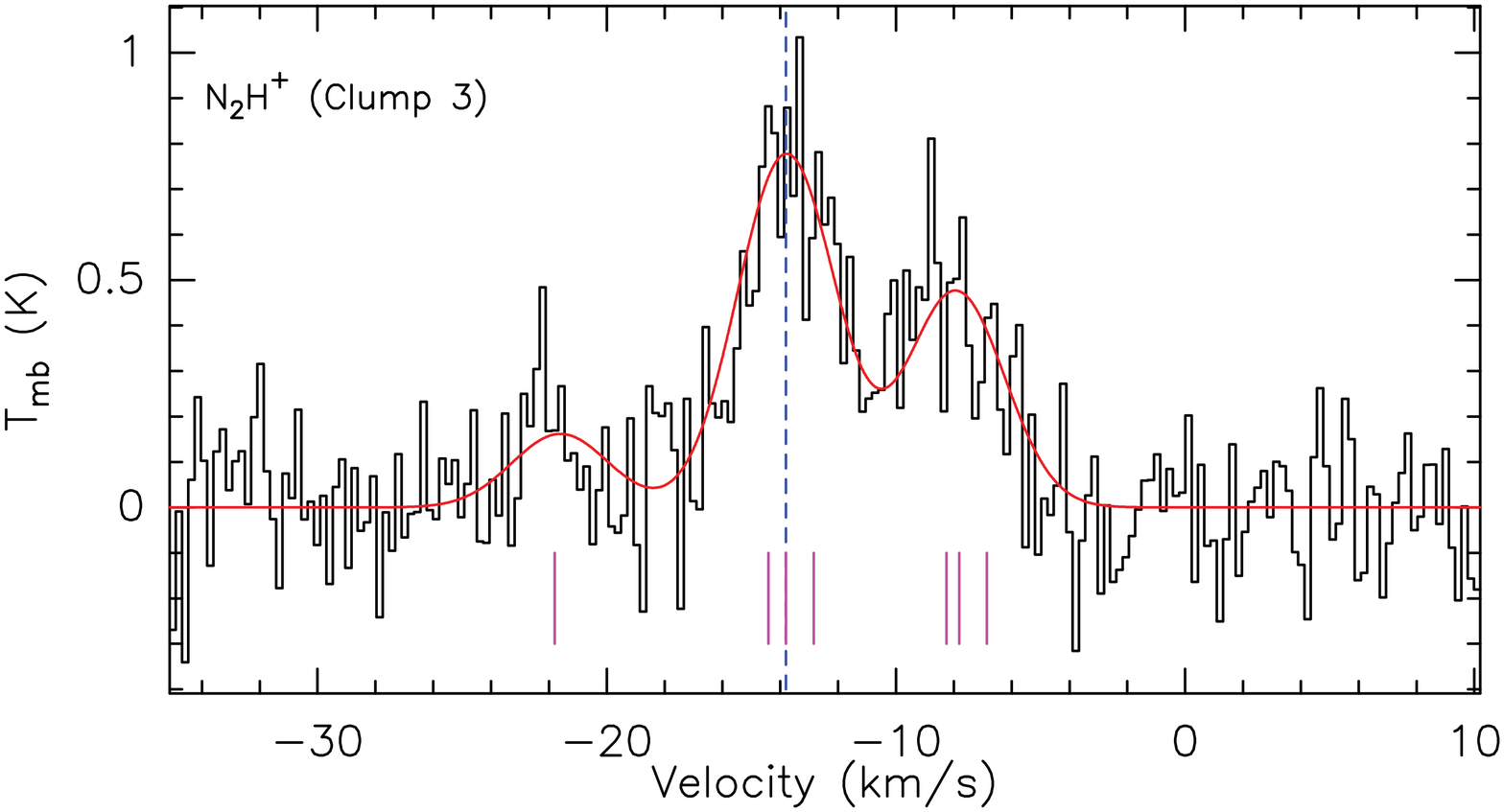}
\caption{Spectra of the optically thin N$_{2}$H$^{+}$ line emission extracted over the three identified clumps associated with IRAS 17149$-$3916. The red curves are the {\tt hfs} fit to the spectra. The estimated $V_{\textup{LSR}}$ is denoted by the dashed blue line and the location of the hyperfine components by magenta lines.}
\label{n2hplus_spectra}
\end{figure*}

We use the optically thin $\rm N_2 H^+$ line emission to determine the $V_{\textup{LSR}}$ and the line width, $=\Delta V$ of the clumps. The line spectra are extracted by integrating over the retrieved apertures of the clumps as shown in Fig.~\ref{n2hplus_spectra}. The $\rm N_2 H^+$ spectra have seven hyperfine structures and the {\tt hfs} method of {\tt CLASS90} is used to fit the observed spectra. The line parameters retrieved from spectra are listed in Table \ref{n2hplus_fit_parameters}. The $V_{\textup{LSR}}$ determined agrees well with the value of $\rm 13.7\, km s^{-1}$ obtained from $\rm CS(2-1)$ observations of the region by \citet{1996A&AS..115...81B} 
\begin{table}
\centering
\caption{The retrieved $\rm N_2 H^+$ line parameters, $V_{\textup{LSR}}$, $\Delta V$, $T_{\rm mb}$ and $\int T_{\textup{mb}} {\rm dV}$ for the identified clumps associated with IRAS 17149$-$3916.}
\begin{tabular}{ccccc}
\hline
Clump & $V_{\textup{LSR}}$ & $\Delta V$    & $T_{\rm mb}$ & $\int T_{\textup{mb}} {\rm dV}$ \\
      & (km s$^{-1}$)      & (km s$^{-1}$) & (K)          & (K km s$^{-1}$)             \\ \hline
1     & -13.0              & 3.2           & 1.5          & 5.4                         \\
2     & -13.8              & 3.7           & 1.6          & 6.8                         \\
3     & -13.8              & 3.9           & 0.8          & 3.1                         \\ \hline
\end{tabular}
\label{n2hplus_fit_parameters}
\end{table}

\section{Discussion}
\label{discussion}
\subsection{Understanding the morphology of the ionized gas}
As seen, the GMRT maps reveal the large extent and prominent cometary morphology of the {\hiir} associated with IRAS 17149$-$3916 which was earlier discussed as a roundish {\hiir} by \citet{2014MNRAS.437..606T}. In this section, we attempt to investigate the likely mechanism for this observed morphology. 
The widely accepted models to explain the formation of cometary {\hiirs} are (1) the bow-shock model (e.g. \citealt{1992ApJ...394..534V}), (2) the champagne-flow model (e.g. \citealt{1979A&A....71...59T}), and (3) the mass loading model (e.g. \citealt{1997ApJ...476..166W}). However, subsequent studies, like those conducted by \citet{2003ApJ...596..344C} and \citet{2006ApJS..165..283A}, find 
the `hybrid' models, that are a combination of these, to better represent the observed morphologies. 

The bow-shock model assumes a wind-blowing, massive star moving supersonically through a dense cloud. Whereas, the champagne-flow model invokes a steep density gradient encountered by the expanding {\hiir} around a newly formed stationary, massive star possibly located at the edge of a clump. Here, the ionized gas flows out towards regions of minimum density. In comparison, the model proposed by \citet{1997ApJ...476..166W} invokes the idea of strong stellar winds mass loading from the clumpy molecular cloud and the cometary structure unfolds when a gradient in the geometrical distribution of mass loading centres are introduced. In this model the massive, young star is considered to be stationary as in the case of the champagne-flow model. 

While observation of ionized gas kinematics is required to understand the origin of the observed morphology, in the discussion that follows, we discuss a few aspects based on the
radio, column density, and FIR maps of the region associated with IRAS 17149$-$3916 along with the identification of E4 as the likely ionizing star (refer Section \ref{ionizing_star}). Following the simple analytic expressions discussed in \citet{2018A&A...612A..36D}, we derive a few shock parameters to probe the bow-shock model. Taking the spectral type of E4 to be O6.5V -- O7V as estimated from the radio flux density, and assuming it to move at a typical speed of $\rm 10~km/s$ through the molecular cloud, we calculate  the `stand-off' distance to range between $\rm 2.6~arcsec (0.02~pc) - 3.1~arcsec (0.03~pc)$. This is defined as the distance from the star at which the shock occurs and where the momentum flux of the stellar wind equals the ram pressure of the surrounding cloud. The theoretically estimated value is significantly less than the observed distance of $\sim \rm 84~arcsec (0.8~pc)$ between E4 and the cometary head. Taking viewing angle into consideration would decrease the theoretical estimate thus widening the disparity further. Based on the above estimations, it is unlikely that the bow-shock model would explain the cometary morphology. To confirm further, we determine the trapping parameter which is the inverse of the ionization fraction. As the ionizing star moves supersonically through the cloud, the swept off dense shells trap the {\hiir} within it and its expansion is eventually inhibited by the ram pressure. Trapping becomes more significant when recombinations far exceed the ionizing photons. Studies of a large number of cometary {\hiirs} show the trapping parameter to be much greater than unity \citep{1991ApJ...369..395M}. For IRAS 17149$-$3916, we estimate the value to lie in the range $3.2 - 3.5$ which indicates either weak or no bow shock. Similar interpretations are presented in \citet{2018A&A...612A..36D} and \citet{2016MNRAS.456.2425V}. The trapping parameters obtained by these authors lie in the range 1.2 -- 4.3.

To investigate the other models, namely the champagne-flow and clumpy/mass loading wind models, we compare the observed spatial distribution of the dust component and the ionized gas. The FIR and column density maps presented in Section \ref{mir-dust} show a complex morphology of pillars, arcs, filaments in the region with detected massive clumps towards the cometary head. The steep density gradient towards the cometary head is evident. Without the ionized and molecular gas kinematics information, it is difficult to invoke the champagne-flow model. However, the maps do show the presence of clumps towards the cometary head which could act as potential mass loading centres and thus support the clumpy cloud model. Further observations and modelling are essential before one can completely understand the mechanisms at work.

\subsection{Ionizing massive star(s)}\label{ionizing_star}
\citet{2006AJ....131..951R} have studied the associated stellar population towards IRAS 17149$-$3916 in the NIR. Using the colour-magnitude diagram, these authors show the presence of a cluster of massive stars within the infrared nebula and suggest IRS-1 to be the likely ionizing source. In a later study, \citet{2014MNRAS.437..606T} have supported this view citing the spectroscopic classification of IRS-1 as O5 -- O6 by \citet{2005A&A...440..121B} and consistency with the Lyman continuum photon flux estimated from the radio observations by  \citet{2013A&A...550A..21S}. 
\begin{figure*}
\centering
\includegraphics[width=9cm,height=6.6cm]{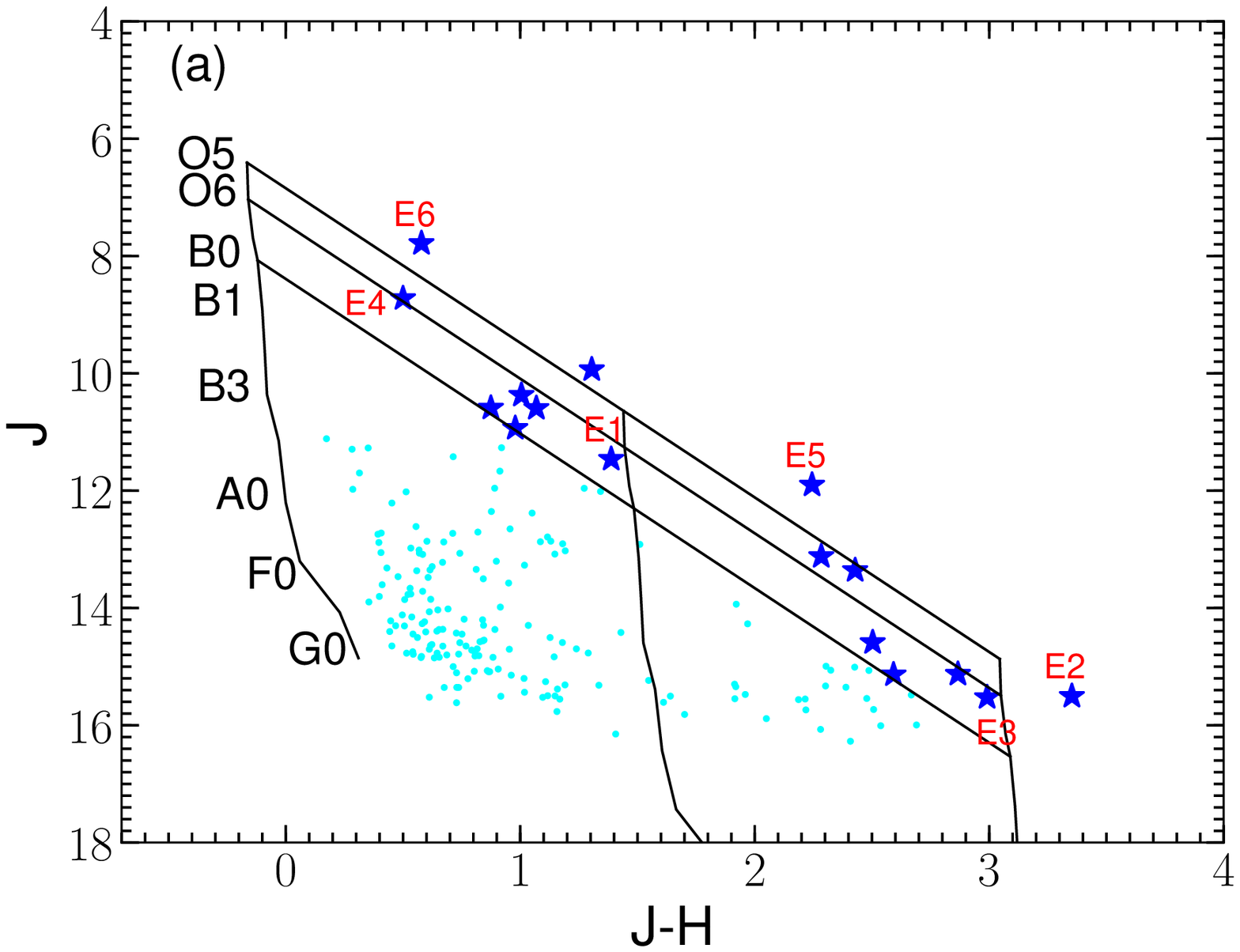}
\includegraphics[width=8cm,height=6cm]{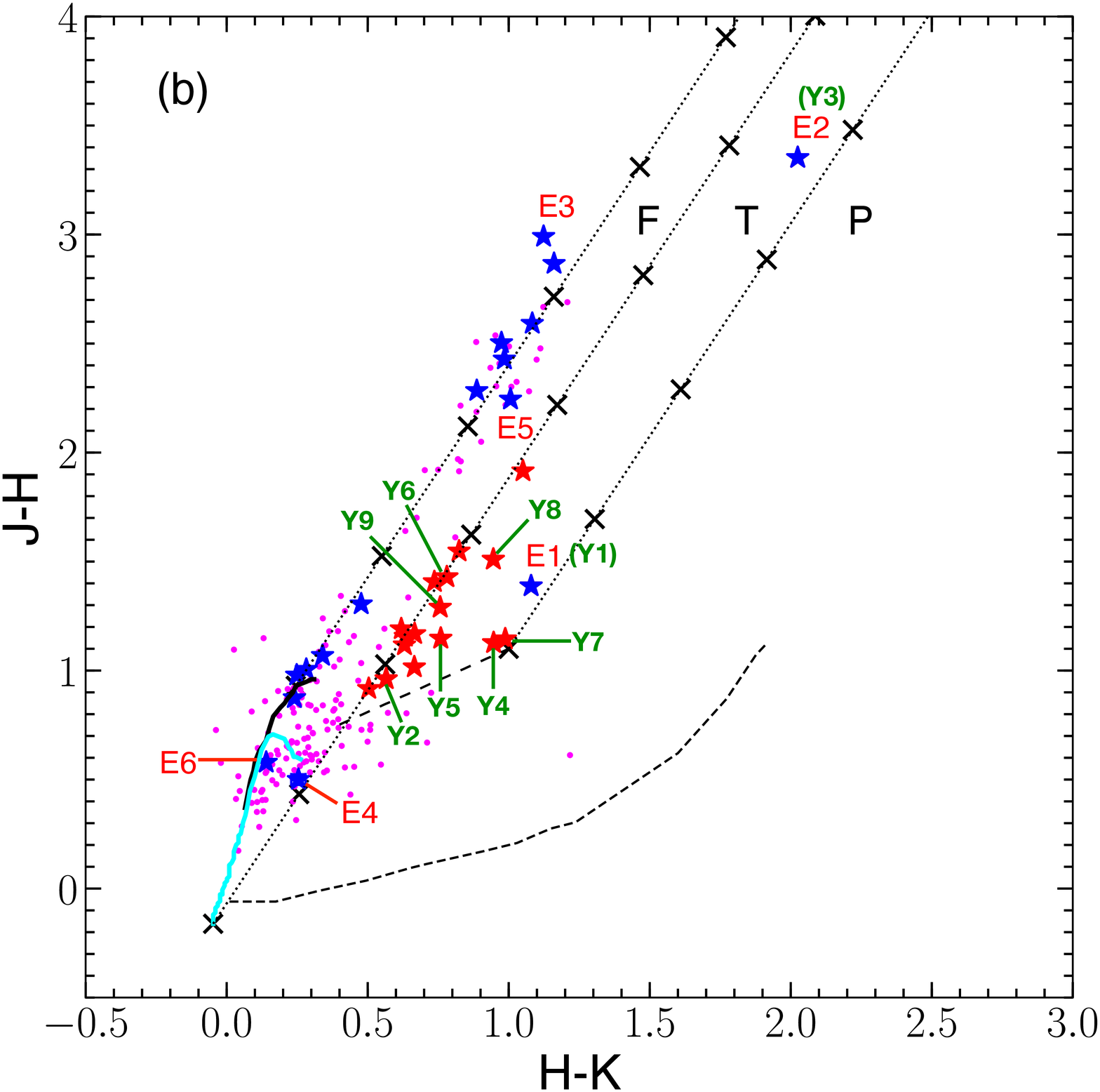}
\includegraphics[width=9cm,height=7cm]{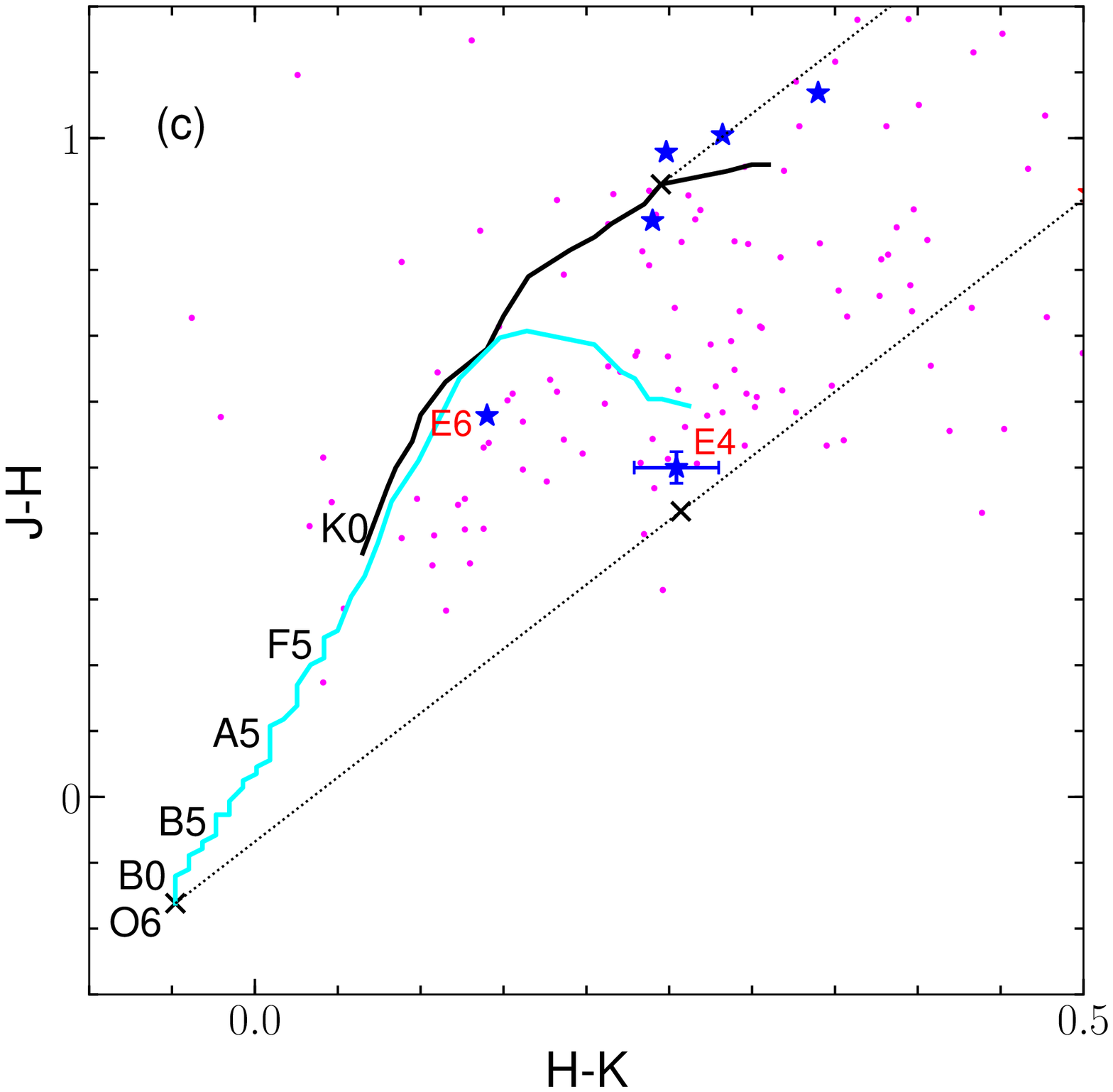}
\caption{(a) J vs J-H colour magnitude diagram of the sources (cyan dots) associated with IRAS 17149$-$3916 and located within the 3$\sigma$ radio contour. The nearly vertical solid lines represent the ZAMS loci with 0, 15, and 30 magnitudes of visual extinction corrected for the distance. The slanting lines show the reddening vectors for spectral types B0 and O5. 
(b) J-H vs H-K colour-colour diagram for sources (magenta dots) in the same region as (a). The cyan and black curves show the loci of main sequence and giants, respectively, and are taken from \citet{1983A&A...128...84K} and \citet{1988PASP..100.1134B}. The locus of classical T Tauri adopted from \citet{1997AJ....114..288M} is shown as long dashed line. The locus of Herbig AeBe stars shown as short dashed line is adopted from \citet{1992ApJ...393..278L}. The parallel lines are the reddening vectors where cross marks indicate intervals of 5 mag of visual extinction. The colour-colour plot is divided into three regions, namely, `F', `T', and `P' (see text for more discussion). The interstellar reddening law assumed is from \citet{1985ApJ...288..618R}. The magnitudes, colours and various loci plotted in both the diagrams are in the \citet{1988PASP..100.1134B} system. 
The identified early type (earlier than B0) and YSOs candidates are highlighted as blue and red stars, respectively.  Of these, the ones located towards the central, bright radio emission are labelled as `E' and `Y', respectively. (c) An enlarged view of the bottom left portion of (b) showing the position of spectral types on the main sequence locus. The location of the source E4 and the errors on the colour are also shown.}
\label{NIR-CC-CM}
\end{figure*}
\begin{figure}
\centering
\includegraphics[scale=0.2]{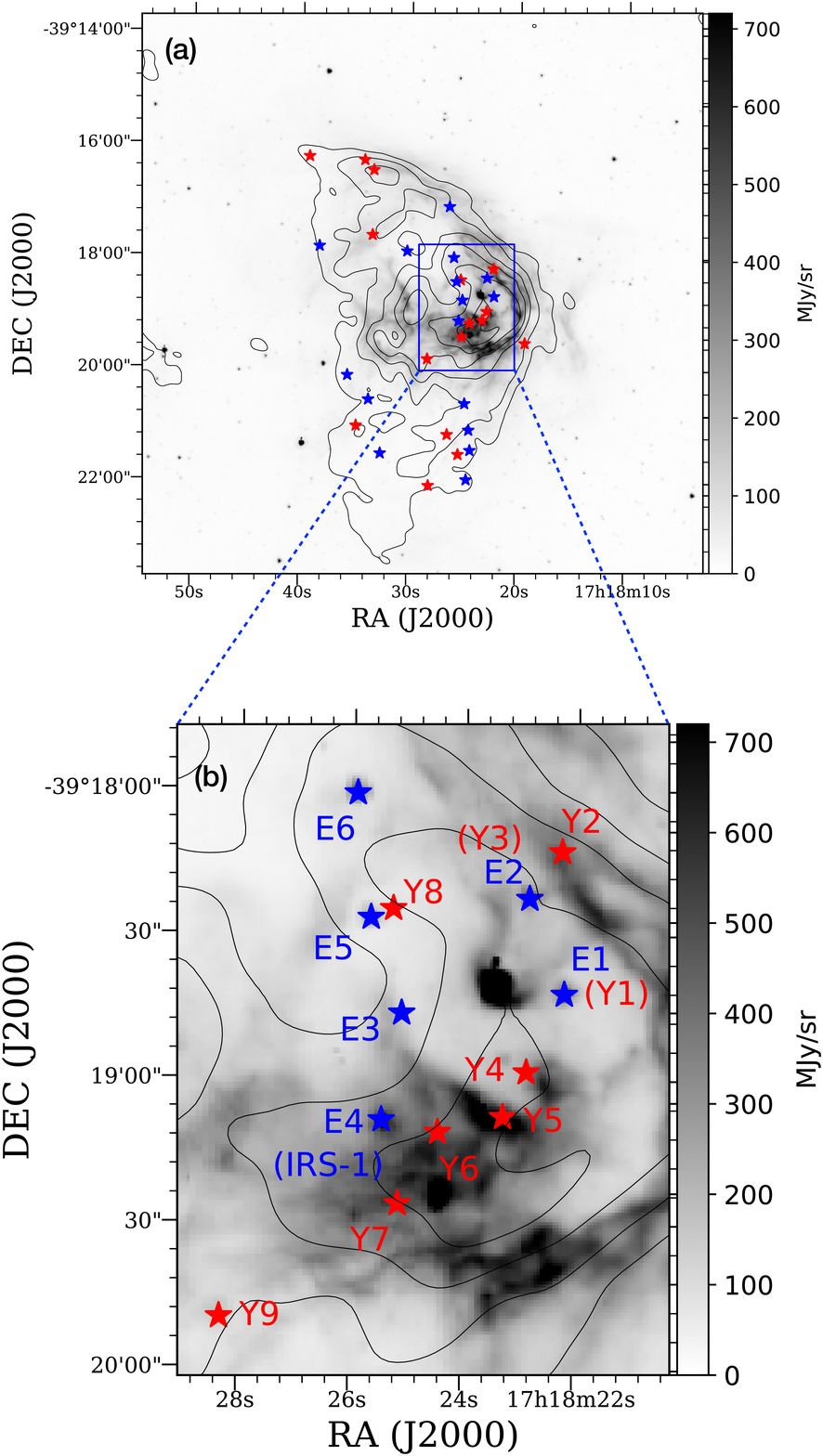}
\caption{Panel a: {\it Spitzer} 5.8\,$\rm \mu m$ image in grey scale overlaid by 610~MHz radio contours. The contour levels are 0.002, 0.01, 0.03, 0.06, 0.12, and 0.2~Jy/beam. The YSOs and massive stars detected towards the central ionized region are shown in red and blue coloured stars, respectively. Panel b: This shows the zoom-in view of central part of the \hii\ region.}
\label{EY-Spa-dis}
\end{figure}
While the spectral type estimated from the GMRT radio emission at 1280~MHz is consistent with that of IRS-1, we investigate the stellar population within the radio emission for a better understanding. 
In Figs.~\ref{NIR-CC-CM}(a) and (b), we plot the NIR colour-magnitude and colour-colour diagrams, respectively, of 2MASS sources located within the 3$\sigma$ radio contour. 
Fig.~\ref{NIR-CC-CM}(c) shows an enlarged view of the bottom left portion of Fig.~\ref{NIR-CC-CM}(b) to highlight the location of the star E4 with respect to the main sequence locus.
Following the discussion given in \citet[Fig. 7;][]{2006A&A...452..203T} the colour-colour plot is classified into `F', `T', and `P' regions. The `F' region is occupied by mostly field stars or Class III sources, the `T' region is for T-Tauri stars (Class II YSOs) and protostars (Class I YSOs) populate the `P' region.
As seen from the figures, there are sixteen sources earlier than spectral type B0 and eighteen identified YSOs. The sample of identified YSOs fall in the `T' region, the sources of which are believed to be Class II objects with NIR excess. Sources that lie towards the central bright, radio emitting region are labelled in the figures with prefixes of `E' for the sources earlier than B0 and `Y' for the YSOs. As indicated in Fig.~\ref{NIR-CC-CM}(b), early type sources E1 and E2 are also the identified YSOs, Y1 and Y3, respectively. The coordinates and NIR magnitudes of these selected sources are listed in Table~\ref{EY-sources}. Fig.~\ref{EY-Spa-dis} shows the spatial distribution of the above sources with respect to the radio and 5.8\,$\rm \mu m$ emission.

As seen from the above analysis, in addition to the presence of possible discrete radio sources that could be internally ionized, several massive stars are also identified from the NIR colour-magnitude and colour-colour diagrams. Hence, it is likely that ionization in this {\hiir} is the result of this cluster of massive stars.  However, the observed symmetrical, cometary morphology of the ionized emission strongly suggests that the ionization is mostly dominated by a single star. 
As seen in Fig.~\ref{NIR-CC-CM}(a), out of the early type sources that lie towards the central, bright radio emission, the colour and magnitude of the source E4 is consistent with a spectral type of $\sim$O6.
A careful scrutiny of the Fig.~\ref{NIR-CC-CM}(b) show that early type stars E1 and E2 are embedded Class II sources and hence unlikely to be the main driving source. Sources E3, E5, E6 are possibly reddened giants or field stars. The location of early type star, E4 (which is the source IRS-1) in the colour-colour diagram (see enlarged view shown in Fig.~\ref{NIR-CC-CM}c) agrees fairly well with the spectral type estimate of $\sim$O6 obtained from the colour-magnitude diagram and strongly advocates it as the dominating exciting source. This is consistent with the identification of IRS-1 as the ionizing star in previous studies. Spatially also, the location of E4 clearly suggests its role in the formation of the network of pillar like structures observed (see Section \ref{pillars}). As mentioned earlier, the spectral type of E4, estimated from NIR spectroscopy, is in good agreement with the radio flux. Location wise, however, it is 30~arcsec away from the radio peak. This offset could be attributed to density inhomogeneity or clumpy structure of the surrounding, ambient ISM. Supporting this scenario of E4 being the dominant player, are the interesting pillar like structures revealed in the MIR images discussed in the next section. 
\begin{table*}
\caption{Early type and YSOs detected within the central, bright, radio emission of IRAS 17149$-$3916}
\begin{center}
\centering
\scalebox{0.85}{
\begin{tabular}{cccccc}
\hline
Source             & \multicolumn{2}{c}{Coordinates} & J      & H      & K      \\
                   & RA (J2000)     & DEC (J2000)    &        &        &        \\ \hline
 &               & {\it Early-type sources}  &        &        &   \\[1mm]
E1 (Y1) & 17 18 22.21   & -39 18 42.24   & 11.364 & 10.039 & 8.957  \\
E2 (Y3) & 17 18 22.85   & -39 18 22.45   & 15.356 & 12.086 & 10.086 \\
E3      & 17 18 25.11   & -39 18 46.41   & 15.398 & 12.486 & 11.361 \\
E4      & 17 18 25.45   & -39 19 08.61   & 8.654  & 8.208  & 7.927  \\
E5      & 17 18 25.68   & -39 18 26.65   & 11.790  & 9.617  & 8.606  \\
E6      & 17 18 25.94   & -39 18 00.89   & 7.715  & 7.191  & 7.021  \\ \hline
  &               &   {\it YSOs}             &        &        &        \\[1mm]
Y1 (E1) & 17 18 22.21   & -39 18 42.24   & 11.364 & 10.039 & 8.957  \\
Y2      & 17 18 22.28   & -39 18 12.69   & 15.064 & 14.161 & 13.578 \\
Y3 (E2) & 17 18 22.85   & -39 18 22.45   & 15.356 & 12.086 & 10.086 \\
Y4      & 17 18 22.87   & -39 18 58.45   & 14.413 & 13.346 & 12.393 \\
Y5      & 17 18 23.28   & -39 19 07.77   & 12.993 & 11.906 & 11.135 \\
Y6      & 17 18 24.44   & -39 19 11.04   & 14.324 & 12.958 & 12.166 \\
Y7      & 17 18 25.14   & -39 19 25.95   & 14.741 & 13.657 & 12.664 \\
Y8      & 17 18 25.28   & -39 18 24.76   & 12.818 & 11.372 & 10.420  \\
Y9      & 17 18 28.30   & -39 19 49.71   & 14.677 & 13.449 & 12.680 \\
\hline
\end{tabular}}
\label{EY-sources}
\end{center}
\end{table*}

\subsection{Triggered star formation}\label{pillars}

\subsubsection*{Pillar Structures}
\begin{figure}
\centering
\includegraphics[scale=0.15]{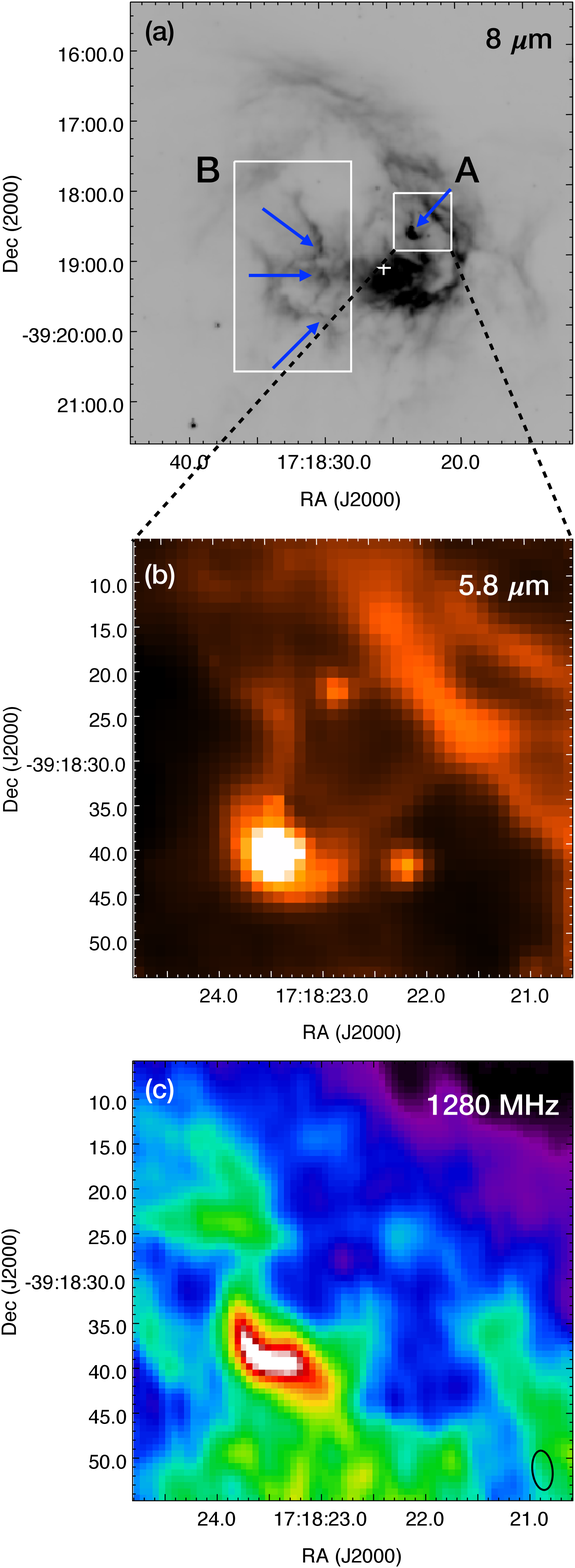}
\caption{(a) 8.0\,$\rm \mu m$ map of IRAS 17149$-$3916 from the {\it Spitzer}-GLIMPSE survey. Blue arrows highlight the pillar structures identified within Boxes `A' and `B'. The white `+' mark shows the position of E4 (IRS-1). (b) A zoom-in on `A' at IRAC 5.8\,$\rm \mu m$. (c) 1280\,MHz map covering the pillar `A'.}
\label{pillar-structure}
\end{figure}

In Fig.~\ref{pillar-structure}, we illustrate the identification of pillar structures in the IRAC 8\,$\rm \mu m$ image. The MIR emission presents a region witnessing a complex interplay of the neutral, ambient ISM with the ionizing radiation of newly formed massive star(s). The boxes labelled `A' and `B' show prominent pillar structures, the orientation of these are clearly pointed towards E4. This strongly suggests E4 as the main sculptor of the detected pillars. Furthermore, it also supports the identification of E4 as the main ionizing source of the \hiir\ . 
One of the mechanisms widely accepted to explain the formation of these pillars is the radiative driven implosion (RDI) \citep{1994A&A...289..559L}. Here, pre-existing clouds exposed to newly forming massive star(s) are sculpted into pillars by slow photoevaporation caused due to strong impingement of ionizing radiation. The other being the classical collect and collapse model of triggered star formation proposed by \citet{1977ApJ...214..725E}. Under this framework, the expanding {\hiir} sweeps up the surrounding material, creating dense structures that could eventually form pillars in their shadows. 

Figs.~\ref{pillar-structure}(b) and (c) show the zoomed in IR and radio view of pillar `A'. Clearly seen is a slightly elongated and bright radio source at the pillar head.  To ascertain the nature of the bright radio source, we estimate few physical parameters using the 1280~MHz GMRT map. Using the 2D fitting tool of {\small CASA} viewer we fit a 2D Gaussian and determine the deconvolved size and flux density of this source to be $\rm 19.05~arcsec \times 8.07~arcsec$ ($\theta_{\rm source} = 12.4 \rm arcsec; 0.12 \rm pc$) and $\rm 445 \,mJy\,$, respectively. Inserting these values in equations \ref{Lyman_flux}, \ref{e_no_density}, and \ref{emission_measure}, we get $ log N_{\rm Ly} = 47.30$ , $n_{\rm e} = 3.9\times10^3~\rm cm^{-3}$ and EM= $\rm 1.8\times10^6~pc~cm^{-6}$. The ionized mass ($\rm M_{ion} = \frac{4}{3}\pi r^{3} \mathit{n_{\rm e}} m_{p}$, where $\rm r$ is the radius of the source and $\rm m_{p}$ is the mass of proton) is also calculated to be $\rm 0.09 ~ M_{\odot}$. 
The estimated values of these physical parameters lie in between the typical values of compact and UCHII region \citep{2002ASPC..267...81K,2005A&A...433..205M,2021A&A...645A.110Y}.
Hence this radio source could well represent an intermediate evolutionary stage between a compact and an UC{\hiir} thus indicating a direct signature of triggered star formation at the tip of the pillar. An alternate picture for the bright radio emission at the head of the pillar could also be external ionization by the ionizing front emanating from E4. Such externally ionized, tadpole-shaped structures have been studied in the Cygnus region, where the ionized front heads point towards the central, massive Cygnus OB2 cluster  \citep{2019A&A...627A..58I}. In support of the former scenario of the UC{\hiir}, a bright and compact 5.8\,$\rm \mu m$ emission region is seen that is co-spatial with radio emission. This compact IR emission is seen in all IRAC bands and {\it Herschel} images. While it is not listed as an IRAC point source in the GLIMPSE catalog, it is included in the PACS 70~$\rm \mu m$ point source catalog \citep{2017arXiv170505693M}. There also exists a 2MASS counterpart (within $\sim$3~arcsec) but has been excluded from the YSO identification procedure owing to poor quality photometry in one or more 2MASS bands. It is thus likely that the compact IR emission sampled in the {\it Spitzer}-GLIMPSE and {\it Herschel} images is the massive YSO powering the UC{\hiir}.

Several studies (e.g. \citealt{2010ApJ...712..797B,2017MNRAS.470.4662P}) have shown evidence of star formation in the pillar tips in the form jets, outflows, YSO population, etc. 
The driving mechanism for this triggered star formation, RDI, is initiated when the propagating ionizing front traverses the pillar head creating a shell (known as the ionized boundary layer, IBL) of ionized gas. If the pressure of the IBL exceeds the internal pressure of the neutral gas within the pillar head then shocks are driven into it. This leads to compression and subsequent collapse of the clump leading to star formation. However, to comment further on the detected UC{\hiir} on the tip of pillar `A' and link its formation to RDI, one needs to conduct pressure balance analysis using molecular line data as discussed in \citet{2017MNRAS.470.4662P} and \citet{2013A&A...556A.105O}. These authors have used $\rm ^{13}CO$ transitions for their analysis which is not possible in our case as CO molecular line data with adequate spatial resolution is not available for this region of interest. Furthermore, attempting any study with the detected MALT90 transitions is difficult given the limited spatial resolution. In a recent study, \citet{2020MNRAS.493.4643M} have carried out involved hydrodynamical simulations to study pillar formation in turbulent clouds. As discussed by these authors, star formation triggered in pillar heads can be explained without invoking the RDI mechanism. Gravitational collapse of pre-existing clumps can lead to star formation without the need for ionizing radiation to play any significant role. From their simulations, they conclude that compressive turbulence driven in {\hiirs}, which competes with the reverse process of photoevaporation of the neutral gas, ultimately dictates the triggering of star formation in these pillars. Further high-resolution studies are required to understand the nature of the compact radio and IR emission at the head of the pillar `A'.

\subsubsection*{Dust clumps and the collect and collapse mechanism}
\label{clumps-CC}
The detection of clumps and the signature of fragmentation to cores is evident from the FIR and sub-mm dust continuum maps presented in Fig.~\ref{dustclumps}. Investigating the collect and collapse hypothesis is necessary to explain whether the dust clumps are a result of swept-up material that is accumulated or these clumps are pre-existing entities. Towards this, we carry out a simple analysis and evaluate few parameters such as the dynamical age ($t_{\rm dyn}$) of the {\hiir} and the fragmentation time ($t_{\rm frag}$) of the cloud. 

Assuming that the {\hiir} associated with IRAS 17149$-$3916 expands in a homogeneous cloud, the dynamical timescale can be estimated using the classical expressions from \citet{1978ppim.book.....S} and \citet{1980pim..book.....D},
\begin{equation}
t_{\rm dyn} = \frac{4}{7}\frac{R_{\rm St}}{a_{\rm Hii}}\left [ \left ( \frac{R_{\rm if}}{R_{\rm St}} \right )^{7/4} - 1  \right ] 
\end{equation}
where, $a_{\rm Hii}$ is the isothermal sound speed and is assumed to be 10~km s$^{-1}$, $R_{\rm if} = 1.8$~pc is the radius of the {\hiir} determined from the geometric mean of an ellipse visually fit to encompass the ionized emission in the 610~MHz GMRT map. $ R_{\rm St}$ is the Str\"{o}mgren radius, given by the following equation
\begin{equation}
R_{\rm St} = \left ( \frac{3 N_{\rm Ly}}{4 \pi n^{2}_{0} \alpha_{\rm B}} \right )^{1/3} 
\end{equation}
where, $N_{\rm Ly}$ is the Lyman continuum flux, $n_{0}$ is the initial particle density of the ambient gas. To derive $n_{0}$, we assume that the dense, bright region seen in the column density map is swept-up material due to the expansion of the \hiir\ and it was initially homogeneously distributed within the radius of the {\hiir}. To estimate the mass of this swept-up material, we use equation \ref{mass_eqn} and the column density map (refer Section \ref{dust_clump}). Integrating within the black contour shown in Fig.~\ref{cdtchi}(a), the mass is estimated to be $1645~ {M_{\odot}}$. Taking the observed estimate of the radius, we calculate $n_{0}$ to be $9.5 \times 10^{2}$ cm$^{-3}$.  $\alpha_{B}$ is the coefficient of radiative recombination and is determined using the expression \citep{1980pim..book.....D}:
\begin{equation}
\alpha_{\rm B} = 2.6\times10^{-13} \left ( \frac{10^{4}\textup{K}}{T_{\rm e}} \right )^{0.7} \textup{cm}^{3} \: \textup{s}^{-1}
\end{equation}
where, $T_{\rm e} = 5000$~K, is the electron temperature. Using the above parameters and from the spectral type of the ionizing source (O6.5V - O7V), we estimate the $t_{\rm dyn}$ to be $\sim$ 0.2~Myr.

Using the formalism discussed in \citet{1994MNRAS.268..291W}, we proceed next to estimate the fragmentation time scale of a cloud and that can be written as
\begin{equation}
t_{\rm frag} = 1.56 \: a^{7/11}_{{\rm s},0.2}\:  \mathit{N^{-1/11}_{{\rm Ly},\textup{49}}}\:  n^{-5/11}_{0,3} \: \textup{Myr}
\end{equation}
where, $a_{\rm s} = a_{\rm s,0.2} \times 0.2$ km s$^{-1}$ is the speed of sound in the shocked layer and is taken as 0.3 km s$^{-1}$ \citep{2017MNRAS.472.4750D}. $N_{\rm Ly} = N_{\rm Ly,\textup{49}} \times 10^{49}$ s$^{-1}$ is the ionizing photon flux and $n_{0} = n_{0,3}\times 10^{3}$ cm$^{-3}$ is the initial particle density of the ambient gas. Plugging in the values in the above expression, we estimate $t_{\rm frag}$ to be $\sim$ 2.2~Myr.
Comparing the estimates of the two time scales involved, it is seen that the fragmentation time scale is more than a factor of 10 larger than the dynamical time scale of the {\hiir}. This essentially indicates that if the clumps detected are the result of swept-up material due to expansion of the {\hiir}, then the shell has not got enough time to fragment thus making the collect and collapse process highly unlikely here. Such a scenario has been invoked by \citet{2012A&A...544A..39J} for the dust bubble N22. In contrast, \citet{2017MNRAS.472.4750D}, in their investigation of bubble CS51 found support for the collect and collapse hypothesis. Thus, further studies, as indicated earlier, are required to probe the RDI process not only with regards to the pillar structures but also the detected clumps. 

\subsection{Nature of the detected dust clumps and cores}
\subsubsection*{Virial analysis of the dust clumps}
Here, we investigate the gravitational stability of the identified dust clumps associated with IRAS 17149$-$3916. This would enable us to determine whether these clumps are gravitationally bound or not. The virial mass, $M_{\rm vir}$, of a dust clump is the amount of mass that can be supported against self-gravity purely by thermal and non-thermal gas motion. This is given by \citet{2016MNRAS.456.2041C}
\begin{equation}
M_{\rm vir} = \frac{5\ r\ \Delta V^2}{8\ {\rm ln}(2)\ a_1\ a_2\ G} \sim 209\ \frac{1}{a_1\ a_2} \left(\frac{\Delta V}{\rm km\ s^{-1}} \right)^2\ \left(\frac{r}{\rm pc}\right) M_{\odot}
\end{equation}
In the above equation, $\Delta V$ is the line width of the optically thin $\rm N_2 H^+$ line, $r$ is the radius of clumps taken from Table \ref{clump-param}, the constant $a_1$ accounts for the correction for power-law density distribution, and is given by $a_1 = (1-p/3)/(1-2p/5)$, for $p< 2.5$ \citep{1992ApJ...395..140B} where we adopt $p=1.8$ \citep{2016MNRAS.456.2041C}. The constant $a_2$ accounts for the shape of the clump which we assume to be spherical and take $a_2$ as 1.
We also calculate the virial parameter, $\alpha_{\rm vir} = M_{\rm vir}/M_{\rm clump}$. The estimated values of $M_{\rm vir}$ and $\alpha_{\rm vir}$ are listed in Table \ref{clump-param}. 
As discussed in \citet{2013ApJ...779..185K} and \citet{2019ApJ...878...10T}, $\alpha_{\rm vir} = 2$ sets a lower threshold for gas motions to prevent collapse in the absence of magnetic field and/or external pressure. The virial parameter estimate for Clump 1 is $ < 2$ indicating that it is gravitationally bound and hence likely to collapse. However, Clump 2 is marginally above this threshold and Clump 3, which shows signature of star formation in the form of an UC{\hiir} has a higher virial parameter value of 4.1. Similar values of $\alpha > 2$ has been observed for protostellar and prestellar dense cores by \citet{2019ApJ...878...10T}. These authors have used the $\rm C^{18}O$ line and discuss the contribution from turbulence as a primary factor that would significantly affect the line width and hence overestimate the virial mass. While turbulence gets dissipated in the densest region of molecular clouds and the $\rm N_2 H^+$ line used here is a dense gas tracer, it is likely that the resolution of the MALT90 survey does not probe the inner dense cores and the observed velocity dispersion is influenced by the outer and more turbulent region. High-resolution molecular line observations are thus essential to probe the nature of the clumps.  

\subsubsection*{Clump fragmentation and the detected cores}
{\it ALMA} 1.4~mm continuum map (Fig.~\ref{dustclumps}) is seen to resolve the identified dust clumps into a string of cores, with masses ranging between $0.3 - 62.3~M_{\odot}$ and radii $\sim 0.01$~pc, thus indicating a scenario of hierarchical fragmentation. If we assume that fragmentation of the clumps is governed by thermal Jeans instability, then the initially homogeneous gas clump has a Jeans length and mass given by \citet{2019ApJ...886..102S}
\begin{equation}
    \lambda_J = \sigma_{\rm th} \left ( \frac{\pi}{G\rho} \right )^{1/2}
\end{equation}
and 
\begin{equation}
    M_J = \frac{4\pi\rho}{3}\left ( \frac{\lambda_J}{2} \right )^3 = \frac{\pi^{5/2}}{6}\frac{\sigma_{\rm th}^3}{\sqrt{G^3\rho}}
\end{equation}
where $\rho$ is the mass density, $G$ the gravitational constant and $\sigma_{\rm th}$ the thermal velocity dispersion (the isothermal sound speed) and is given by
\begin{equation}
    \sigma_{\rm th} = \left ( \frac{k_B T}{\mu m_{\rm H}} \right )^{1/2}
\end{equation}
where $ k_B$ is the Boltzmann constant and $\mu$ the mean molecular weight. 
As the thermal velocity dispersion will be dominated by $\rm H_2$ and He, we consider $\mu = 2.37$ \citep{{2008A&A...487..993K},{2014MNRAS.439.3275W},{2019ApJ...886..102S}}. Using the clump parameters tabulated in Table \ref{clump-param}, we estimate $\lambda_J$ and $M_J$ of the clumps which are listed in Table \ref{clump-Jeans}. If turbulence drives the fragmentation instead, then the turbulent Jeans length and mass for each clump is derived by replacing the thermal velocity dispersion with the clump velocity dispersion estimated from the observed line width of the dense gas tracer $\rm N_2 H^+$ which is a good approximation for the turbulent line width. The calculated $\lambda_{\rm turb}$ and $M_{\rm turb}$ values are given in Table \ref{clump-Jeans}. 
\begin{table*}
\caption{Hierarchical fragmentation of clumps associated with IRAS$-$3916}
\begin{center}
\centering
\begin{tabular}{ccccccccc}
\hline
Clump & $\sigma_{\rm th}$ & $\sigma^a$ & $\lambda_J$ & $M_J$ &$\lambda_{\rm turb}$ & $M_{\rm turb}$  \\
      & ($\rm km\,s^{-1}$) & ( $\rm km\,s^{-1}$) & (pc) & ($M_{\odot}$)  & (pc)  & ($M_{\odot}$) \\  
\hline 

1   & 0.3   & 1.4   & 0.2   & 6.8   & 0.8   & 543 \\
2   & 0.3   & 1.6   & 0.2   & 5.3   & 0.9   & 807 \\
3   & 0.3   & 1.7   & 0.1   & 5.6   & 0.8   & 819\\

\hline
\end{tabular}
\label{clump-Jeans}
\end{center}
$^a$ $\sigma = \Delta V/\sqrt{8\,{\rm ln}2}; \Delta V$ being the $\rm N_2 H^+$ line width.
\end{table*}
The turbulent Jeans masses are $\sim 80 - 150$ times larger than the thermal Jeans mass. 
Comparing with the derived core masses, it is seen that 11 out of the 15 detected cores ($\sim 73\%$) in the three clumps have masses less than the Jeans mass. This suggests that the observed cores are consistent with the prediction of Jeans fragmentation without invoking turbulence indicating that it does not play a significant role in the fragmentation process. Similar results are obtained by \citet{2019ApJ...886..102S} who studied the 70\,$\rm \mu m$ dark massive clumps in early stages using {\it ALMA} data. As discussed by these authors, the majority of detected cores having masses less than the thermal Jeans mass supports competitive accretion and hierarchical fragmentation frameworks. The four cores whose mass exceeds the Jeans mass (the `super-Jeans' cores) are suitable candidates for forming high-mass cores. However, further high-resolution observations are essential to completely understand the fragmentation process, if any, at the core level. 

\begin{figure*}
\centering 
\includegraphics[scale=0.6]{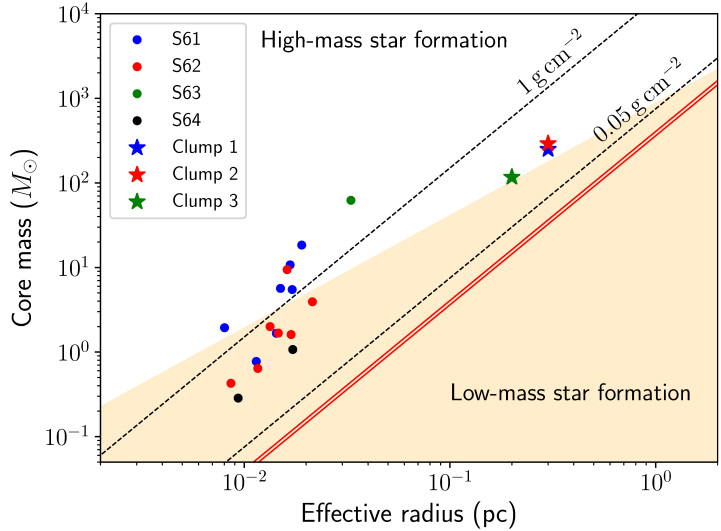}
\caption{Masses of the dense cores identified from the 1.4\,mm {\it ALMA} maps and the cold dust clumps identified using the 350\,$\rm \mu m$ \textit{Herschel} map of IRAS 17149$-$3916 are plotted as a function of their effective radii, depicted by circles and `$\star$'s, respectively. The shaded area corresponds to the low-mass star-forming region that do not satisfy the condition, $ M > 870\,M_\odot(r/\rm pc)^{1.33}$ \citep{2010ApJ...723L...7K}. Black-dashed lines indicate the surface density thresholds of 0.05 and 1\,$\rm g\,cm^{-2}$ defined by \citet{2014MNRAS.443.1555U}  and \citet{2008Natur.451.1082K}, respectively. The red lines represent the surface density thresholds of 116\,$M_\odot\,\rm pc^{-2}$ ($\sim 0.024\,\rm g\,cm^{-2}$) and 129\,$M_\odot\,\rm pc^{-2}$ ($\sim 0.027\,\rm g\,cm^{-2}$) for active star formation proposed by \citet{2010ApJ...724..687L} and \citet{2010ApJ...723.1019H}, respectively.}
\label{alma_cores_mass_radius}
\end{figure*}
In Fig.~\ref{alma_cores_mass_radius}, we plot the estimated mass and radius of the identified clumps and cores. The plot also compiles several surface density thresholds proposed by various studies to identify clumps/cores with efficient and active star formation  \citep{2010ApJ...724..687L,2010ApJ...723.1019H,2014MNRAS.443.1555U}. In addition, criteria for these to qualify as high-mass star-forming ones are also included. All the detected clumps and cores associated with IRAS 17149$-$3916 are seen to be active star-forming regions. The three clumps satisfy the empirical mass-radius criteria, $ M > 870\,M_\odot(r/\rm pc)^{1.33}$ defined by \citet{2010ApJ...723L...7K}, and hence are likely to harbour massive star formation. At the core scale ($\rm < 0.1~pc$), \citet{2008Natur.451.1082K} have posed a theoretical surface density threshold of $\rm 1\,g\,cm^{-2}$, below which cores would be devoid of high-mass star formation. From the figure, we see that there are four cores (2 in Clump 1, 1 each in Clumps 2 and 3) which have masses $\gtrsim 10~M_{\odot}$ and above this surface density limit. These are the `super-Jeans' cores discussed above. 
High-resolution molecular line observations are essential to shed better light on the nature of the cores, the gas kinematics involved and for accurate determination of physical parameters like temperature, mass, etc.

\subsection{Conclusion}
\label{conclusion}
Using multiwavelength data, we have carried out a detailed analysis of the region associated with IRAS 17149$-$3916. The important results of this study are summarized below.

\begin{enumerate}
\item Using the GMRT, we present the first low-frequency radio continuum maps of the region mapped at 610 and 1280~MHz. The {\hiir}, previously believed to be nearly spherical, displays a large-extent cometary morphology. The origin of this morphology is not explained by the bow shock model. The presence of dense clumps towards the cometary head indicates either the champagne flow or the clumpy cloud model but further observations of the ionized gas kinematics are essential to understand the observed morphology. 

\item The integrated flux densities yields an average spectral index value of $-0.17\pm0.19$ consistent with thermal {\it free-free} emission. If powered by a single massive star, the estimated Lyman continuum photon flux suggests an exciting star of spectral type  O6.5V -- O7V star. 

\item NIR colour-magnitude and colour-colour diagrams show the presence of a cluster of massive stars (earlier than spectral type B0) located within the bright, central radio emitting region. MIR and FIR images show complex and interesting features like a bubble, pillars, clumps, filaments,and arcs revealing the profound radiative and mechanical feedback of massive stars on the surrounding ISM. 

\item The spatial location of source, E4 (IRS-1), and the orientation of observed pillar structures with respect to it, strongly suggest it as the dominant driving source for the cometary {\hiir}. This view finds support from the position of E4 in the colour-magnitude and colour-colour diagrams. Further, its spectral type estimation from literature agrees well with that estimated for the exciting source of the {\hiir} from GMRT data. 

\item The column density map reveals the presence of dust clumps towards the cometary head while the dust temperature map appears to be relatively patchy with regions of higher temperature within the radio nebula. The dust clumps identified using the  \textit{Herschel} 350\,$ \rm \mu m$ map have masses ranging between $\sim$100 - 300~$\rm M_\odot$ and radius $\sim$0.2 - 0.3~pc. Virial analysis using the $\rm N_2 H^+$ shows that the south-east clump (\#1) is gravitationally bound. For the other two clumps (\# 2 and 3), the line widths would possibly have contribution from turbulence thus rendering larger values of the virial parameter.

\item A likely compact/UC{\hiir} is seen at the tip of a pillar structure oriented towards the source E4 thus suggesting evidence of triggered star formation under the RDI framework. In addition, the detected dust clumps are investigated to probe the collect and collapse model of triggered star formation. The estimated dynamical time scales are seen to be smaller by a factor of $\sim$10 compared to the fragmentation timescale of the clumps thus clearly negating the collect and collapse mechanism at work. 

\item The {\it ALMA} 1.4~mm dust continuum map probes the dust clumps at higher resolution and reveal the presence of 17 compact dust cores with masses and radii in the range of $0.3 - 62.3~M_\odot$ and 0.01 -- 0.03~pc, respectively. The largest and the most massive core is located within Clump 3. The estimated core masses are consistent with thermal Jeans fragmentation and support the competitive accretion and hierarchical fragmentation scenario. 

\item Four `super-Jeans' fragments are detected and are suitable candidates for forming high-mass stars and their mass and radius estimates satisfy the various threshold defined in literature for the potential high-mass star-forming cores.

\end{enumerate}

\section*{Acknowledgements}
We would like to thank the referee for comments and suggestions which helped in improving the quality of the manuscript. We thank the staff of the GMRT that made the radio observations possible. GMRT is run by the National Centre for Radio Astrophysics
of the Tata Institute of Fundamental Research. The authors would like to thank Dr. Alvaro S\'{a}nchez-Monge for providing the FITS image of the radio maps presented in \citet{2013A&A...550A..21S}. CHIC acknowledges the support of the Department of Atomic Energy, Government of India, under the project  12-R\&D-TFR-5.02-0700. This work is based
[in part] on observations made with the {\it Spitzer} Space Telescope, which is operated by the Jet Propulsion Laboratory, California Institute of Technology under a contract with NASA. This publication also made use of data products from {\it Herschel} (ESA space observatory). This publication makes use of data products from the Two Micron All Sky Survey, which is a joint project of the University of Massachusetts and the Infrared Processing and Analysis Center/California Institute of Technology, funded by the NASA and the NSF. This work makes use of the ATLASGAL data, which is a collaboration between the Max-Planck-Gesellschaft, the European Southern Observatory (ESO) and the Universidad de Chile. This paper makes use of the following ALMA data: ADS/JAO.ALMA\#2016.1.00191.S. ALMA is a partnership of ESO (representing its member states), NSF (USA) and NINS (Japan), together with NRC (Canada), MOST and ASIAA (Taiwan), and KASI (Republic of Korea), in cooperation with the Republic of Chile. The Joint ALMA Observatory is operated by ESO, AUI/NRAO and NAOJ. This research has made use of the SIMBAD database, operated at CDS, Strasbourg, France. 

\addcontentsline{toc}{section}{Acknowledgements}
\section*{Data Availability}
The original data underlying this article will be shared on reasonable request to the corresponding author.


\bibliographystyle{mnras}
\bibliography{refer} 


\bsp	
\label{lastpage}
\end{document}